\documentclass[sigconf]{acmart}
\AtBeginDocument{%
  \providecommand\BibTeX{{%
    \normalfont B\kern-0.5em{\scshape i\kern-0.25em b}\kern-0.8em\TeX}}}

\newcommand{\userquote}[1]{\textit{``#1''}}

\usepackage{booktabs}  
\usepackage{amsmath}  
\usepackage{tabularx}
\newcolumntype{R}{>{\raggedleft\arraybackslash}X} 

\usepackage{graphicx}
\usepackage{subcaption}

\usepackage{listings}
\usepackage{pifont}

\usepackage{xcolor, soul}
\definecolor{neonfuchsia}{rgb}{1.0, 0.25, 0.39}

\definecolor{blue}{RGB}{0,0,0}
\newcommand{\bluehighlight}[1]{{\textcolor{blue}{#1}}}

\setcopyright{acmcopyright}
\copyrightyear{2024}
\acmYear{2024}
\acmDOI{XXXXXXX.XXXXXXX}

\acmConference[CHI '24]{}{}{Honolulu, HI}
\acmBooktitle{CHI 2024} 
\acmPrice{15.00}
\acmISBN{978-1-4503-XXXX-X/18/06}

\copyrightyear{2024}
\acmYear{2024}
\setcopyright{acmlicensed}\acmConference[CHI '24]{Proceedings of the CHI Conference on Human Factors in Computing Systems}{May 11--16, 2024}{Honolulu, HI, USA}
\acmBooktitle{Proceedings of the CHI Conference on Human Factors in Computing Systems (CHI '24), May 11--16, 2024, Honolulu, HI, USA}
\acmDOI{10.1145/3613904.3642698}
\acmISBN{979-8-4007-0330-0/24/05}

\begin{document}

\title{\bluehighlight{\textit{CoQuest}: Exploring Research Question Co-Creation with an LLM-based Agent}}

\author{Yiren Liu}
\email{yirenl2@illinois.edu}
\affiliation{%
  \institution{University of Illinois Urbana-Champaign}
  \country{USA}
}

\author{Si Chen}
\email{sic3@illinois.edu}
\affiliation{%
  \institution{University of Illinois Urbana-Champaign}
  \country{USA}
}

\author{Haocong Cheng}
\email{haocong2@illinois.edu}
\affiliation{%
  \institution{University of Illinois Urbana-Champaign}
  \country{USA}
}

\author{Mengxia Yu}
\email{myu2@nd.edu}
\affiliation{%
  \institution{University of Notre Dame}
  \country{USA}
}

\author{Xiao Ran}
\email{xiaor2@illinois.edu}
\affiliation{%
  \institution{University of Illinois Urbana-Champaign}
  \country{USA}
}

\author{Andrew Mo}
\email{jiajunm3@illinois.edu}
\affiliation{%
  \institution{University of Illinois Urbana-Champaign}
  \country{USA}
}

\author{Yiliu Tang}
\email{yiliut2@illinois.edu}
\affiliation{%
  \institution{University of Illinois Urbana-Champaign}
  \country{USA}
}

\author{Yun Huang}
\email{yunhuang@illinois.edu}
\affiliation{%
  \institution{University of Illinois Urbana-Champaign}
  \country{USA}
}



\begin{abstract}

Developing novel research questions (RQs) often requires extensive literature reviews, especially in interdisciplinary fields. 
To support RQ development through human-AI co-creation, we leveraged Large Language Models (LLMs) to build an LLM-based agent system named \textit{CoQuest}. 
We conducted an experiment with 20 HCI researchers to examine the impact of two interaction designs: breadth-first and depth-first RQ generation. The findings revealed that participants perceived the breadth-first approach as more creative and trustworthy upon task completion. Conversely, during the task, participants considered the depth-first generated RQs as more creative. Additionally, we discovered that AI processing delays allowed users to reflect on multiple RQs simultaneously, leading to a higher quantity of generated RQs and an enhanced sense of control. Our work makes both theoretical and practical contributions by proposing and evaluating a mental model for human-AI co-creation of RQs. 
We also address potential ethical issues, such as biases and over-reliance on AI, advocating for using the system to improve human research creativity rather than automating scientific inquiry. The system's source is available at: \href{https://github.com/yiren-liu/coquest}{https://github.com/yiren-liu/coquest}.



\end{abstract}

\ccsdesc[500]{Human-centered computing~Empirical studies in HCI}
\ccsdesc[500]{Human-centered computing~Interactive systems and tools}
\ccsdesc[500]{Computing methodologies~Natural language processing}
\keywords{Scientifc Discovery, Large Language Models, Co-creation Systems, Mixed-initiative Design}





\begin{teaserfigure}
    \includegraphics[width=\textwidth]{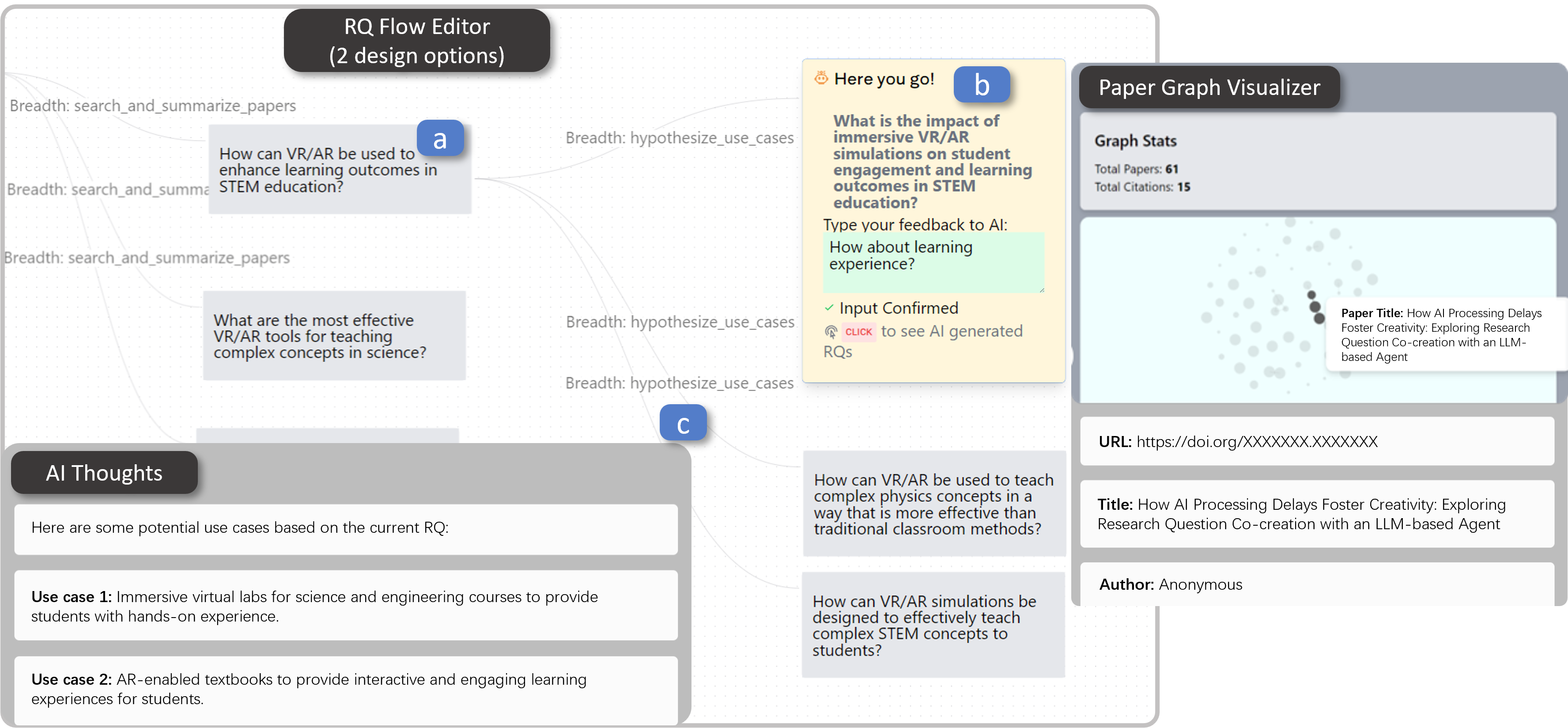}
    \caption{
      \bluehighlight{\textit{CoQuest} enables Human-AI co-creation of research questions (RQs) using LLMs through a three-panel design: \textit{RQ Flow Editor}, \textit{Paper Graph Visualizer}, and \textit{AI Thoughts}. Major features of the \textit{RQ Flow Editor} panel include: (a) offering either breadth-first and depth-first RQ generation (see Figure \ref{fig:2conditions}); (b) enabling users to provide feedback to AI based on a generated RQ, and to explore related papers in details on the \textit{Paper Graph Visualizer} panel; and (c) presenting the rationale behind RQ generation on the \textit{AI Thoughts} panel when a link between two RQs is clicked. }
    }
    \label{fig:system_interface_overview}
    \Description{This is an overview of the CoQuest system interface. There are three panels. The main panel is the RQ Flow Editor Panel with 2 design options, detailed in Figure 4. There are six nodes presented on this panel: three on the left, three on the right. The three nodes on the left are each connected to a node not included in this diagram. The top left node is connected to three nodes on the right. Each edge had text associated on it. The top right node is expanded with more content and a different color. The top left node is denoted (a), the expanded node on the top right is denoted (b), and one of the edges is denoted (c). On the right of RQ Flow Editor Panel is the Paper Graph Visualizer Panel, which presents an interactive graph. Each paper node on the network represents a paper. One paper node is selected and the details of this paper is presented below the graph on this panel. On the bottom-left of the RQ Flow Editor Panel is the AI Thoughts Panel, which has three blocks of texts on it.}
\end{teaserfigure}

\maketitle

\section{Introduction}


Identifying research questions (RQs) is a critical step of scientific research and discovery \cite{elio2011computing}. 
To formulate creative and valuable research ideas, researchers often start with searching and scoping the relevant literature, especially when the search involves works spanning multiple domains \cite{Kang2022AugmentingSC}.
Developing research questions is also iterative, that researchers would first have an initial idea or topic in mind, then conduct a literature search to refine the ideas, and repeat this process until they have a satisfying research idea \cite{foster2004nonlinear}. Reading a large body of literature, synthesizing them, and identifying the relevant work can be rather time-consuming.


With the rapid development of Large Language Models (LLMs), scholars have investigated the potential of harnessing the power of LLMs to support the process of research literature discovery \cite{elicit,Qureshi2023AreCA,Consensus2023}, which helps users significantly speed up the literature discovery process. 
Meanwhile, the thriving of generative AI technologies has also been widely used in various fields to promote creativity \cite{Epstein2023ArtAT,yuan2022wordcraft,mirowski2023co}. For example, a recent study has found that more than 4.9\% of users used ChatGPT to support creative ideation \cite{Sparktoro2023}. 
Although recent research has demonstrated the potential of using smaller language models to generate novel RQs \cite{Liu2023CreativeRQ}, there remains a lack of empirical understanding about how humans evaluate AI-generated RQs. 
Given that LLMs are known to have problems with hallucination and lack of factual accuracy \cite{ji2023survey}, creating high-quality RQs requires inputs from human researchers for their unique backgrounds and expertise. Enabling humans and AI to co-create novel research questions is particularly promising for conducting interdisciplinary research.

The concept of human-AI co-creation draws insights from pioneering research on mixed-initiative systems, where human users and computer systems contribute to a shared goal  \cite{horvitz1999principles,yannakakis2014mixed}. Recent academic endeavors have also introduced a variety of design guidelines for human-AI co-creation systems \cite{rezwana2022designing, weisz2023toward, lin2023beyond}, further inspiring the intricate design considerations of mixed-initiative systems. For instance, \citet{rezwana2022designing} explored whether the exchange of initiatives between humans and AI should be designed as either parallel or in a turn-taking manner. 
However, there remains a gap in empirical research regarding the influence of AI-driven initiatives or creativity on user experiences, such as their perception of the creative process, trust in the AI, and sense of control. Moreover, the potential impact of these factors on the results of human-AI co-creation, such as the quality of generated content, has yet to be investigated.

To this end, we proposed a novel system called \textit{CoQuest}, which allows an AI agent to initiate RQ generation by tapping the power of LLMs and taking humans' feedback into a co-creation process. The system consists of three panels: \textit{RQ Flow Editor} for supporting RQ generation, \textit{Paper Graph Visualizer} for exploration of literature space, and \textit{AI Thoughts} for explaining AI's rationales.
The system design was informed by a formative study, where we invited four researchers to verbalize their expected RQ-generation processes with AI support. 
The formative study yielded an initial mental model of human-AI co-creation for RQs. The system design was further evaluated with 20 participants \bluehighlight{who are HCI researchers} through a within-subject experiment.

Our work makes novel and significant contributions as follows. Firstly, we made a theoretical contribution by proposing and evaluating a new mental model for human-AI co-creation of research questions in the HCI research domain.
Second, we proposed a new agent LLM system and implemented two interaction designs (breadth-first and depth-first), where the AI agent took different levels of initiative in supporting users to develop RQs. 
Third, through a within-subject experiment with HCI researchers, we gained new empirical understandings of how AI's different initiatives impact users' perceived experiences and outcomes. Specifically, for overall experience, the breadth-first design made users ``feel'' more creative and gained more trust from users, though the effect varies by users' research background; but when rating individual RQs, users scored the depth-first questions for higher creativity. 
Fourth, by closely observing participants' interaction with AI, we discovered important co-creation behavior and proposed a new design implication, namely, intentionally ``slowing down AI", giving wait times for users to explore the co-creation space. This is especially beneficial for users to gain a stronger sense of control.  
Last but not least, we discussed the implications of our study for designing ethical human-AI co-creation systems, along with potential ethical concerns. 
We advocate for employing LLM-based systems to augment human creativity and support the ideation and scaffolding process, rather than using LLM-generated ideas for automating HCI research.


\section{Related Work}

\subsection{Human-Led Literature Discovery and Research Idea Creation}

Scholars have sought to understand the process of how researchers conduct literature discovery and formulate new research ideas \cite{Glueck1975SourcesOR,pertsas2017scholarly}.
Extensive works have strived to formulate the model of researchers' scientific activities \cite{palmer2009scholarly,Vilar2015InformationBO,benardou2010conceptual}. For example, 
\citet{foster2004nonlinear} proposed a framework of idea formulation in academic research from an information behavior perspective, which consists of three major components: Opening (initial exploration), Orientation (problem definition and literature survey) and Consolidation (knowledge creation). In this framework, literature discovery and idea formulation are discussed as a recursive and iterative process. Many studies have also been conducted to understand how researchers produce innovative ideas for research purposes \cite{Yang2016CreativeCA}. 
For example, \citet{Jing2015DevelopingAR} proposed a system for research idea creation that incorporates knowledge reuse through ontology construction. 
Later work by \cite{Guo2020TopicBasedEA} introduced a system for supporting research ideation through topic modeling and visualization. Recently, 
\citet{Liu2023CreativeRQ} proposed to generate new research questions using generative language models fine-tuned over related works and explicitly written research questions from publications within the HCI domain. 
But none of these works leveraged large language models (LLMs) or built an LLM-based system to examine human-AI co-creation RQ processes.

\subsection{Agent-based Large Language Model (LLM) Systems}
Recent surge in the success of LLMs \cite{gpt4, touvron2023llama} has spurred strong interests in employing LLMs for solving complex tasks.
There has recently been a heated wave of explorations in building autonomous agents using LLMs as a reasoning engine to solve different tasks, such as software development \cite{shinn2023reflexion, qian2023communicative}, gaming \cite{wang2023voyager}, and assisting social science research \cite{ziems2023can}.

System designs are proposed for improving the method of prompting in Human-AI interaction systems enabled by LLMs \cite{wu2022ai,wu2022promptchainer,dang2022prompt}.
To enhance the reasoning capabilities of LLMs, researchers proposed several prompting frameworks to elicit reasoning in LLMs by decomposing and solving the sub-tasks step by step by applying prompting techniques for general purposes, such as Chain-of-Thought \cite{wei2022chain, kojima2022large}, Self-Consistency \cite{wang2023selfconsistency}, Least-to-Most \cite{zhou2023leasttomost}. These prompting techniques focus on improving the task-specific performance of LLMs. 
To build autonomous agents with LLMs, \citet{Yao2022ReActSR} proposed a prompting framework named ``ReAct'' that unifies the ability of LLMs to reason, take actions, and observe the results. 
Recent research on LLM-based prompt engineering \cite{wang2023metacognitive} has explored a viable approach of introducing humans' mental model as prompting techniques to boost LLM's reasoning ability.
These frameworks set a great foundation for Q\&A reasoning or general-purpose tasks \cite{Liu2023AgentBenchEL}. In this paper, we applied LLM to build an agent for specialized tasks in the context of literature discovery and research ideation. 

\subsection{Designing Human-AI Co-creation Systems}
Many human-AI co-creation systems utilize generative AI technologies \cite{sun2022investigating,louie2020novice}. For example, \citet{Bilgram2023AcceleratingIW} showed that generative AI could augment the early phases of innovation, such as exploration, ideation, and digital prototyping.
\citet{Epstein2022WhenHA} discovered that discrepancies between users' expected outputs and the actual output from the system play a critical role in creating new ideas.

Scholars and practitioners try to develop design guidelines for using generative AI \cite{weisz2023toward,davis2015enactive,liu2022design,muller2020mixed}.
From a role-based perspective, human-AI co-creation systems are found to be different from prior systems where machines mainly provide support to humans \cite{kantosalo2021role}. 
In co-creation systems, both humans and AIs can be designed to take the initiative in producing creative artifacts \cite{oh2018lead,guzdial2019friend}. 
\citet{rezwana2022designing} introduced a Co-Creative Framework for Interaction design (COFI), which categorizes mixed-initiative system designs into two types based on their styles of participation: parallel and turn-taking.  
The evaluation of mixed-initiative designs has been explored by recent research \cite{kreminski2022evaluating, withington2023right} in application domains such as poem writing and game design. 
Our research seeks to systematically model the co-creation process and provide both theoretical and practical implications in the scholarly research domain.


In this study, we propose and examine a new agent LLM system that helps researchers formulate research questions by combining LLMs' reasoning ability with the mental model we discovered through a formative study with actual researchers. 
We also discuss whether LLM-based co-creation systems can help facilitate the process of information gathering and idea evolution using user study results to provide new empirical understandings.

\section{Research Questions}
Given the above literature and identified research gaps, we aim to address the following research questions:
\begin{enumerate}
    \item[]  \textbf{RQ1 (perception \& outcome):} \textit{How do users \underline{perceive} the co-creation experience (e.g., trust, control, feeling of being creative) and outcomes (creativity ratings of generated RQs) when using the \textit{CoQuest} system?}
    \item[]  \textbf{RQ2 (behavior):} \textit{How do users \underline{interact} with AI when it provides different levels of initiatives during the co-creation process?}
    \item[] \textbf{RQ3 (relationship):} \textit{What behavioral factors are \underline{associated} with users' enhanced perceived experience and outcomes in human-AI co-creation?}
\end{enumerate}

In the remainder of this paper, we will first present the formative study, where we developed an initial mental model for RQ co-creation between humans and AI. We then introduce the system design and our experimental study and detail the findings in the order of the proposed research questions. We will conclude with an in-depth discussion on the updated mental model informed by our findings and explore design implications for future work. 
Note that we use RQ1, RQ2, and RQ3 to refer to our three research questions in the remainder of the paper, whereas RQ and RQs are used to refer to the research questions co-created by the user and AI.

\section{Formative Study}
We conducted a formative study in order to understand the cognitive process of researchers creating research questions. 
\bluehighlight{
To understand how researchers create RQs, we first conducted semi-structured interviews with 4 HCI researchers about how they formulate research questions.
}
We then organized a focus group with the same group of interviewed participants to identify their needs for an RQ co-creation system. 

We invited four researchers to participate in our semi-structured interviews to understand their process of conducting research. All four participants were doctoral researchers. 
We asked the interviewees to describe their most recent experience starting a research project from scratch. 
Most participants mentioned starting from a rough idea (e.g., domains, keywords, application scenarios) before searching for related literature. 
Participants also emphasized that formulating research ideas is often an iterative process. Typically, participants used one or multiple hypothetical research questions/ideas to facilitate the search for related literature and identify research gaps. 
This process was described by the participants as both time-consuming and labor-intensive. 
Participants also identified their process of research question creation as hierarchical. 
One participant explicitly mentioned that creating research questions naturally resembles the form of a ``mind map'', where the development of ideas gradually ``narrows down'' but could have different branches. 
As described by the participants, the initial set of RQs can often be broader and more general and can subsequently derive sub-RQs under the topics of their predecessors, which helps facilitate the trade-off between specific and general ideas as a common decision-making step during the research ideation process.
This process often resulted in the evolution of RQs that, when visualized, could be drawn as a tree of RQs. This later informed the design of our system to take the form of an interactive ``mind map'' that preserved the provenance of RQ development. 







\begin{figure*}[h]
    \centering
    
    \begin{subfigure}{0.4\textwidth} 
        \centering
        \includegraphics[width=\textwidth]{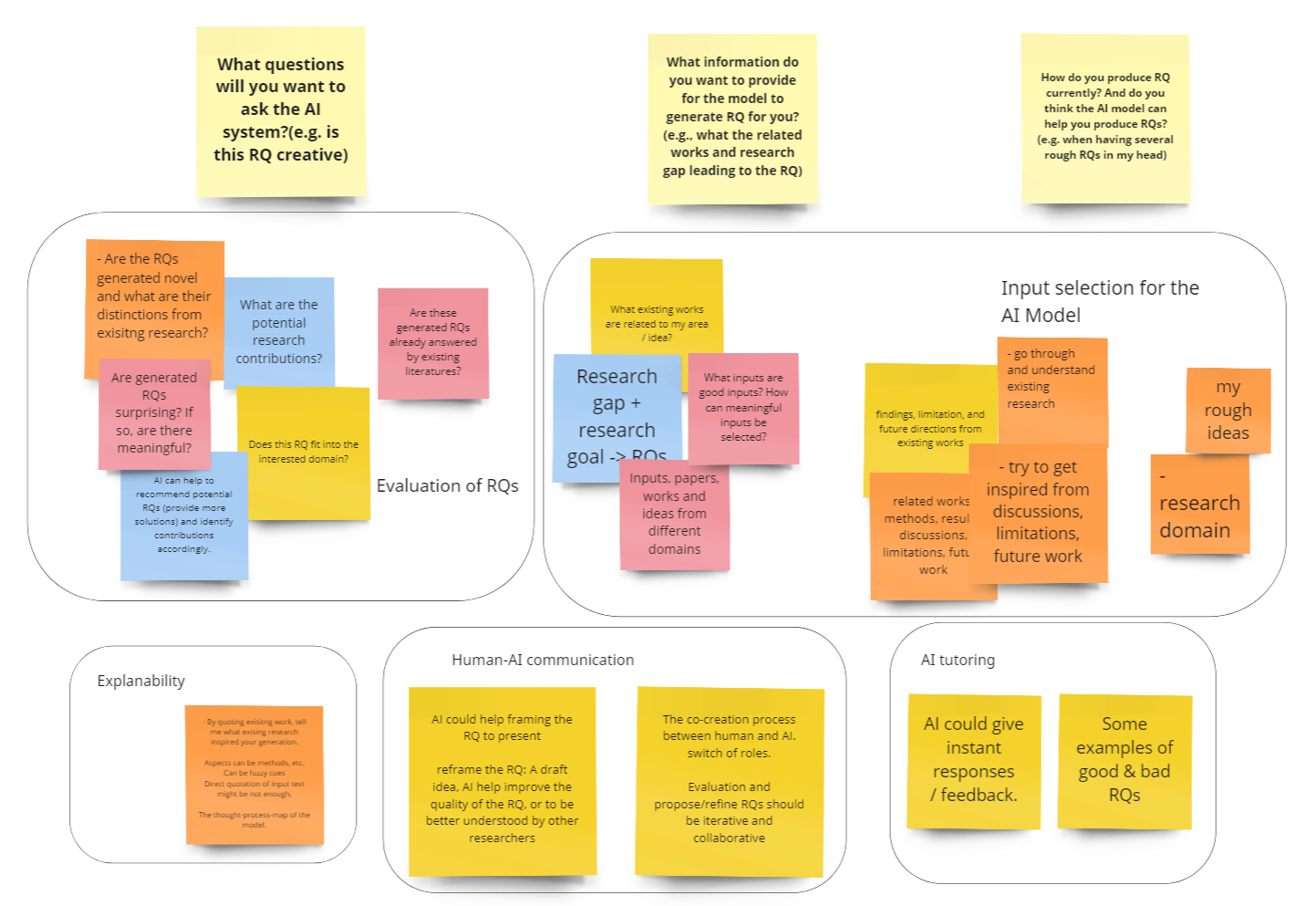}
        \caption{Feature Design.}
        \label{fig:focus_group_1}
        \Description{The feature design screenshot was presented on sticky notes on this diagram. The sticky notes were grouped into five themes, with different number of notes in each group. There are also three yellow notes on top of the groups.}
    \end{subfigure}
    \hfill  
    \begin{subfigure}{0.55\textwidth} 
        \centering
        \includegraphics[width=\textwidth]{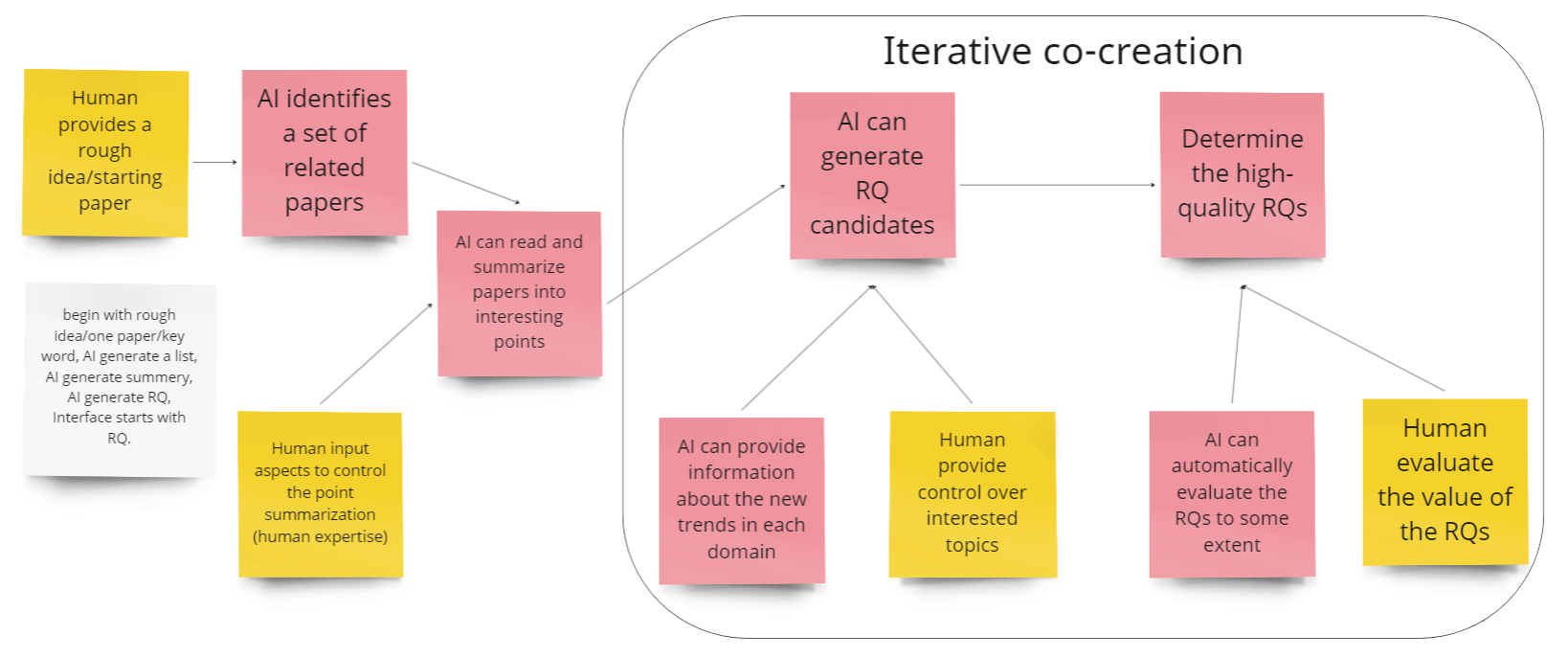}
        \caption{Workflow Design.}
        \label{fig:focus_group_2}
        \Description{The workflow design screenshot was presented on sticky notes on this diagram. Six sticky notes on the right are inside a rectangular shape with text "Iterative co-creation." Five more sticky notes are on the left outside the rectangular. All notes, except one white note on the left, are connected to at least one other note.}
    \end{subfigure}
    \caption{Screenshots of content created by participants using \textit{Miro} during the focus group study.}
    \label{fig:focus_groups}
\end{figure*}

After the first interview study, the same participants were invited back to a focus group session, where we aimed to identify their needs based on their research workflow and propose interaction design to support the process. 
The focus group went through two steps. First, participants were given three questions created by researchers as cues, and then were asked to brainstorm ideas and design expectations given the context of an AI-based system that supports research question development. 
The three question cues were designed as follows: 1) What questions will you ask the AI system? 2) What information do you want to provide for the model to generate RQ for you? 3) How do you produce RQ currently? And do you think the AI model can help you produce RQs?
The ideas created by participants are shown in Figure \ref{fig:focus_groups}.

After discussion and summarizing themes, participants were asked to proceed to the second step inspired by participatory approaches, where they discussed a hypothetical workflow by contextualizing themselves using an AI-based RQ co-creation system.
During the focus group, we found that participants highlighted several expectations for designing a human-AI RQ co-creation system. Participants mentioned that the system should enable strong user control by taking into consideration users' inputs (e.g., ideas, keywords, and domain concepts), when generating new RQs.
The ability for the system to generate different variations of RQs with high diversity was also deemed a preferred design, where the user should be allowed to choose from the outputs based on their preference and expertise.


\textbf{Human-AI Co-Creation of Research Questions (RQs): a Mental Model}
\label{sec:mental_model}
After completing the formative study, we summarized the findings and proposed a new mental model aiming to capture the major interactions during the process of Human-AI Co-Creation for RQs. 


\begin{figure*}[h]
    \centering
    \includegraphics[width=0.9\textwidth]{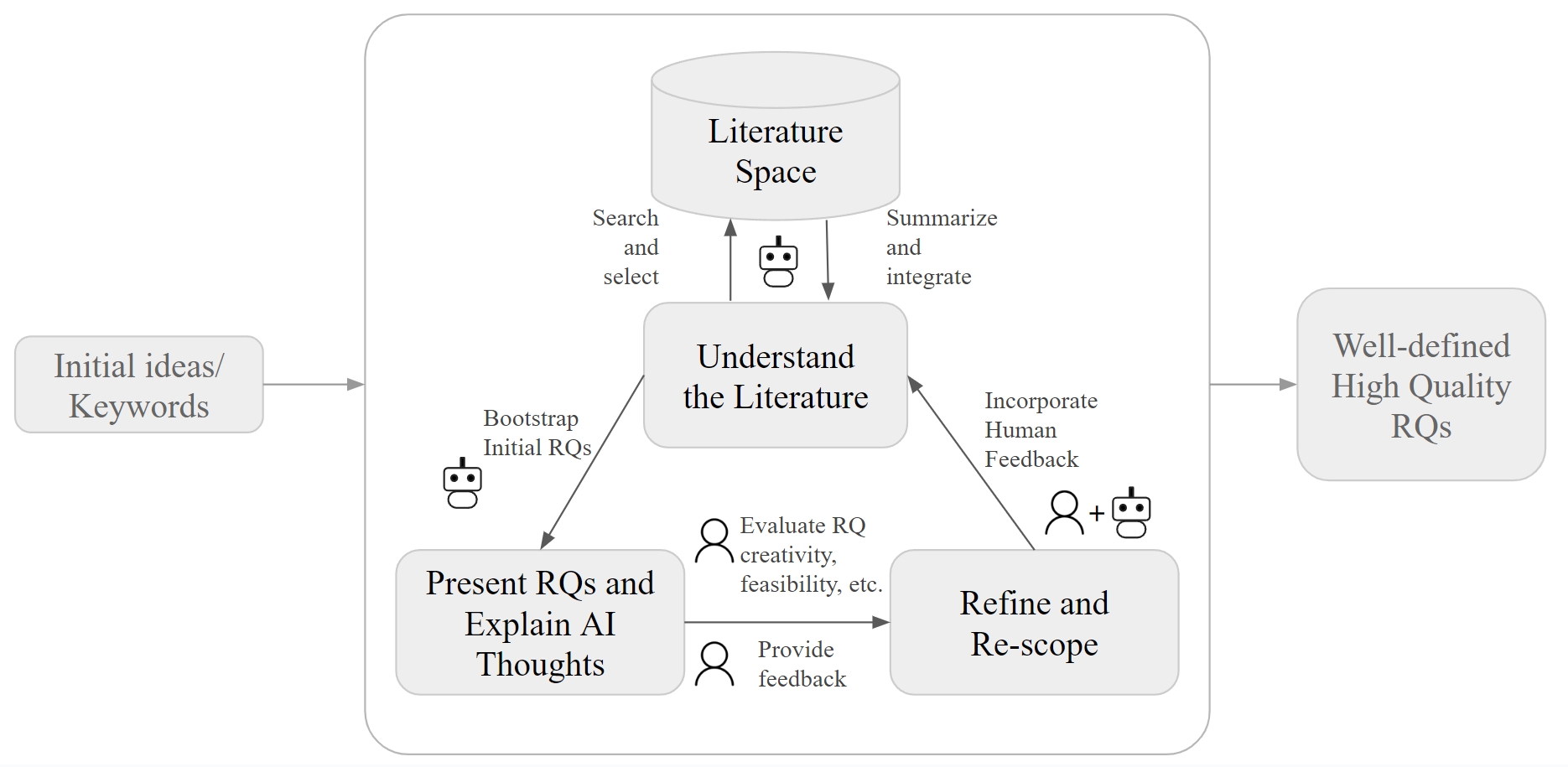}
    \caption{
    Participants' mental model of co-creating RQs with an LLM-based AI agent is delineated as follows. An ``action'' labeled with an AI icon denotes that participants perceived that AI could significantly reduce the ''labor''; a human icon means that participants were expected to evaluate AI-generated RQs and provide feedback to drive the iterative process. However, it is hard to determine how human and AI share the task of ``refine and re-scope,'' as it depends on individual expertise and the clarity of the intended research focus.  
    }
    \label{fig:mental_model}
    \Description{This is human's mental model of RQ co-creation with the LLM-based agent. There are four components inside this model: Present RQs and Explain AI Thoughts (lower left, rectangular), Understand the Literature (center, rectangular), Literature Space (top, with cylinder shape), and Refine and Re-scope (lower right, rectangular). From Present RQs and Explain AI Thoughts to Refine and Re-scope, there are two user-driven activities: Evaluate RQ creativity, feasibility, etc., and Provide feedback. From Refine and Re-scope to Understand the Literature, there is one user- and AI-driven activity: Incorporate Human Feedback. From Understand the Literature to Present RQs and Explain AI Thoughts, there is one AI-driven activity: Bootstrap Initial RQs. From Understand the Literature to Literature Space, there is one AI-driven activity: Search and select. From Literature space to Understand the Literature, there is one AI-driven activity: Summarize and Integrate. The model has "Initial idea / keywords" as input and "Well-defined High Quality RQs" as output.}
\end{figure*}

The proposed mental model consists of three major components: \textit{Understand Literature}, \textit{Present RQs and Explain AI Thoughts}, and \textit{Refine and Re-scope}. 
Past research has provided in-depth discussion over how literature discovery plays a major role in the scientific research process \cite{foster2004nonlinear,palmer2009scholarly}. 
However, such models were often discussed in a human-only context. Our formative study results indicate the importance of considering AI as a ``collaborator'' during the process of research ideation \cite{kim2023effect}. 
Refining research questions has also often been interpreted as a sub-step of literature search and understanding.
Although research ideation and literature search are mutually dependent, our formative study results indicated a distinction between the behaviors researchers conducted during the two different stages. 
The process of literature discovery often involves reading and summarizing a wide range of existing works. Participants emphasized their challenges during the process of literature search and discovery, especially for unfamiliar domains, and the need for AI to facilitate this process with less human involvement. AI can be best used to perform factual summarization and distillation of findings and knowledge before presenting them to humans.
On the contrary, proposing and refining research questions often requires more creative thinking and generalization beyond past knowledge. In this scenario, it is crucial to involve both humans and AI through the design of mix-initiative co-creation systems in order to combine humans' expertise and preference with AI's general world knowledge. 
Moreover, our formative study findings explicate the evaluation of RQs as an additional component during the RQ co-creation process. This process, as discussed during the formative study, should utilize human expertise and the ability to conduct follow-up research and validation. 

\vspace{2mm}

\textbf{Design Requirements}
Based on our findings from the focus group and the proposed mental model, we also propose the following design requirements for the system: The system should be able to 1) assist users' brainstorming process by automatically generating RQ candidates by taking human feedback; 2) support users' sensemaking of AI's outputs by Explaining the rationale behind the generation; 3) help users discover relevant literature and identify research gaps.

\section{\textit{C\MakeLowercase{o}Q\MakeLowercase{uest} }System Design and Implementation}
Based on our findings from the formative study, we designed and implemented an LLM-based system that supports human-AI co-creation of creative research questions. 
In this section, we provide details about the design of 1) the three-panel interface of the \textit{CoQuest} system, including two different designs to provide varied degrees of AI initiative; and 2) the agent LLM backend of the \textit{CoQuest} system.


\subsection{\textit{CoQuest} Interaction Design}

To support RQ co-creation, we designed features of the \textit{CoQuest} systems around the two-way communication between users and the AI, where users and AI take turns during the communication process.
As shown in Figure \ref{fig:system_interface_overview}, our proposed system consists of three major panels: 
\begin{enumerate}
    \item \textit{RQ Flow Editor} 
    that facilitates a user's major interactions, such as generating RQs, providing input and feedback to AI, and editing the RQ flow (e.g., drag and delete);
    \item \textit{Paper Graph Visualizer} 
    that displays the literature space related to each RQ;
    \item \textit{AI Thoughts} 
    that explains AI's rationale of why each RQ is generated.
\end{enumerate}



\subsubsection{\bluehighlight{Example User Walkthrough}}
\bluehighlight{
Consider a user of the \textit{CoQuest} system, Jamie, who is a junior doctoral student with a research direction in Human-Computer Interaction. Jamie has previously been familiar with publications related to interaction design for online learning systems. With recent exposure to social media discussions on VR and AR applications, they wanted to explore the potential of such applications in the domain of online learning. 
Jamie formed an initial idea of using AR to promote learner’s brainstorming. However, without a deeper understanding of the literature space from the VR and AR domain, they found it challenging to refine and improve upon the idea further.
}

\bluehighlight{
Introduced with the \textit{CoQuest} system, Jamie first created an initial idea node, typed down  ``Using AR to promote brainstorming,'' and generated follow-up RQs by right clicking the node and selecting ``generate RQs'' using the \textit{RQ Flow Editor}. 
Aware of Jamie’s initial idea, the \textit{CoQuest} system retrieved several works related to the idea and generated follow-up RQ nodes along with rationales. 
Jamie first saw one of the RQ nodes with an RQ displayed as ``How can social AR be designed to promote collaborative brainstorming?''
}

\bluehighlight{
Jamie was intrigued by the generated RQ, but also a bit confused since the concept of ``social AR'' appeared new to them. Jamie then clicked the RQ node and skimmed through the papers displayed on the \textit{Paper Graph Visualizer} panel. Several papers seemed relevant to Jamie, and they further clicked the provided URL to read the papers in detail. The papers retrieved by \textit{CoQuest} helped Jamie comprehend the domain knowledge behind the generated RQ. 
}

\bluehighlight{
After the reading, Jamie had a question in mind --- ``Why is using AR important for collaborative brainstorming?''. They then typed in this question as user feedback to the system and clicked to generate new RQs following up the previous RQ. One of the newly generated RQs, ``How can spatial design in AR promote group-based collaborative brainstorming,'' caught Jamie’s attention. 
}

\bluehighlight{
The appearance of the concept ``spatial design'' raised Jamie’s interest, but they were unclear about the system’s rationale for generating this RQ from the previous question. By clicking on the edge connecting this RQ with its predecessor, they were able to examine the \textit{AI Thoughts} panel explaining the rationales and actions taken by the agent LLM in the \textit{CoQuest} backend that led to the RQ. The panel showed that the system performed action ``hypothesize user cases'' and one of the resulting use cases was shown as  ``Online learners can form groups and organize ideas by creating and organizing concepts and links in spatial AR environment.''
}

\subsubsection{RQ Flow Editor: Two Design Options} 
\textit{CoQuest} offers an interactive \textit{RQ Flow Editor} panel that allows users to co-create RQs with AI in an iterative manner. 
This panel is designed in a way to resemble the design of a mind map, where each RQ node 
represents a generated RQ (except the initial nodes, which only contain users' initial ideas). 
This allows users to organize RQs more easily under different topics while preserving the hierarchical structure among different RQs, as suggested by the findings from the formative study in section \ref{sec:mental_model}. 
Users can type in their rough ideas or keywords by creating an initial node. 
When users click on one of the nodes, the node expands, and users can 1) type in textual user feedback to AI in the text box; and 2) right-click on the node to generate more follow-up RQs. 
The generated RQs will be connected with annotated edges to the source RQ node.
The RQ generation will result in one or several Directed Acyclic Graphs (DAGs), which we will later refer to as RQ flows, that embed both the hierarchical relations between generated RQs. Users can perform basic interactions using the RQ flow editor, including zooming in/out and dragging the nodes to organize the flows to aid their thinking process.

\begin{figure*}[!t]
    \centering
    \includegraphics[width=0.8\textwidth]{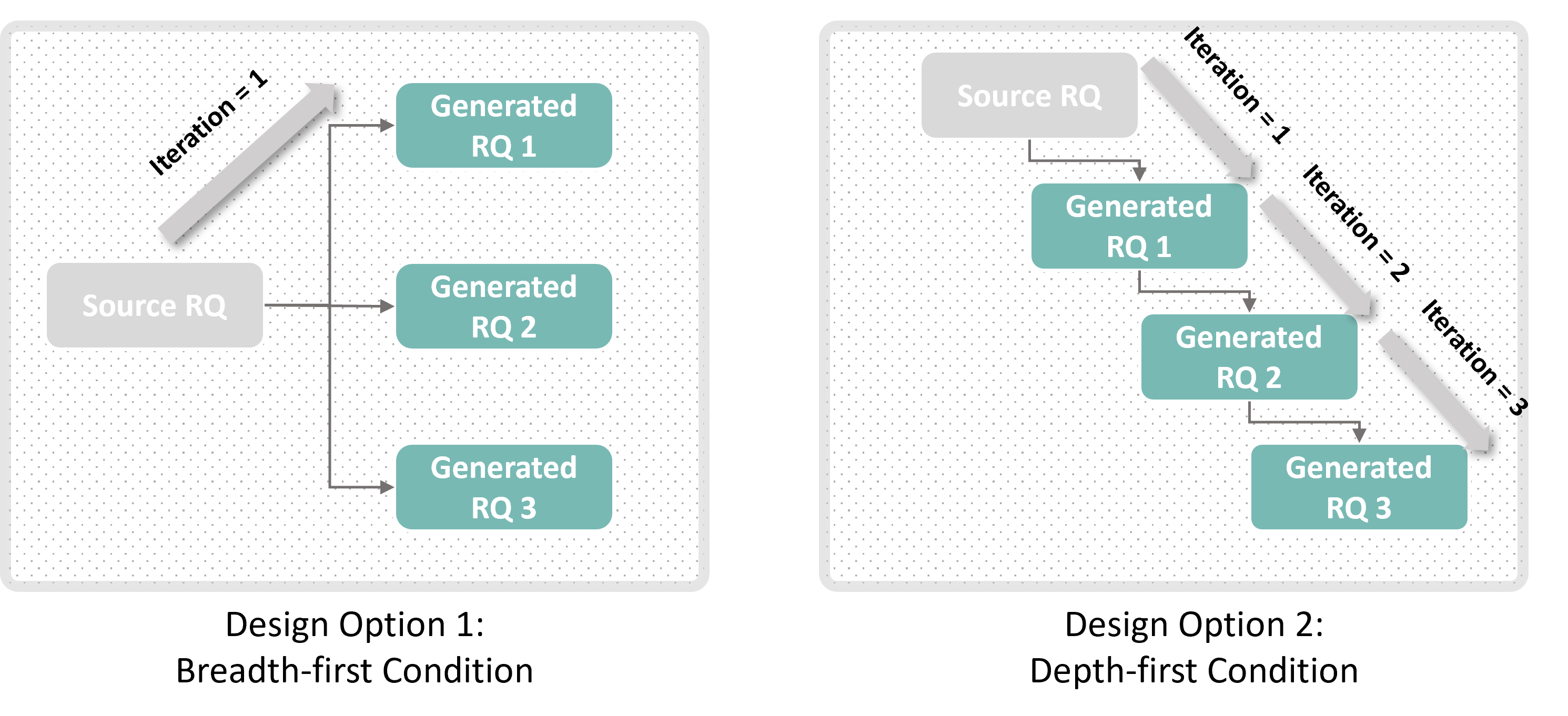}
    \caption{
   The \textit{RQ Flow Editor} panel in the \textit{CoQuest} system features two distinct designs for generating research questions (RQs): the breadth-first and depth-first approaches. The breadth-first generation approach is designed to trigger the creation of multiple RQs in a single iteration, facilitating a wide exploration of potential research areas. In contrast, the depth-first generation focuses on triggering more iterations of RQ refinement, allowing the AI to delve deeper into a specific topic for a more focused and detailed exploration. 
    }
    \label{fig:2conditions}
    \Description{The two design options are presented side by side. On the left is Design Option 1: Breadth-first Generation. From a Source RQ node on the left, three generated RQ nodes on the right were connected with arrows. An additional arrow pointing toward three Generated RQ nodes shows that these three arrows are under iteration 1. On the right is Design Option 2: Depth-first Generation. From a Source RQ node, a first Generated RQ node is connected with an arrow. An additional arrow pointing toward the first Generated RQ node shows that this is iteration 1. From the first Generated RQ, a second Generated RQ node is connected with an arrow. An additional arrow pointing toward the second Generated RQ node shows that this is iteration 2. From the second Generated RQ, a third Generated RQ node is connected with an arrow. An additional arrow pointing toward the third Generated RQ node shows that this is iteration 3.}
\end{figure*}

\textbf{Two design options with different levels of AI initiative: \textit{Breadth-first} and \textit{Depth-first} generation.}
One of the major design options we considered for the \textit{CoQuest} system is the degree of how much AI takes initiative during the co-creation process. 
During the formative study, we obtained an understanding of how users formulated their research questions in a hierarchical form, where follow-up RQs are iteratively created based on previous predecessor RQs. 
Intuitively, when formulating different RQs, the researcher might choose to explore different topics in a broader sense, or to dive deep into a specific topic.
Thus, we consider two different designs when the system generates new RQs, as shown in Figure \ref{fig:2conditions}: 
1) \textit{breadth-first generation}: When a user initiates the generation of follow-up RQs, several RQs are produced \textbf{simultaneously in parallel}. These new questions are all at the same hierarchical level following the original question.
2) \textit{depth-first generation}: In contrast, with this method, when a user initiates the generation, follow-up RQs are created one after the other, with each new question building upon the previous one in a \textbf{sequential} manner.

The two designs impact the degree of initiative taken by AI in the \textit{CoQuest} system. 
Under the breadth-first generation, the user will have the freedom to choose from multiple generated RQs and provide feedback at each turn of generation, and then proceed to generate more RQs if desired. 
As in the depth-first generation, the agent recursively generates a sequence of multiple RQs without user input during the process. 
During the depth-first generation, the agent needs to autonomously further ``refine'' the RQs multiple steps based on the previous steps' results, thus taking more initiative during the co-creation.
Although in both designs, AI actively participates in the co-creation process by turn-taking with the human \cite{rezwana2022designing}, there exists a difference between the degree of initiative taken by AI, as AI engages more actively during the depth-first generation compared to the breadth-first generation.
By carrying out the within-subject study under the two designs, we aim to provide an empirical understanding of how initiative-driven design options can impact users' behavior and perception using human-AI co-creation systems. 
To simplify the study design, we ensured that three RQs were generated per turn in both designs.


\subsubsection{Paper Graph Visualizer: Interactive Literature Graph} 
The \textit{CoQuest} system also provides an LLM-enabled literature discovery feature to assist users in efficiently identifying and exploring existing works related to each generated RQ. 
As shown in Figure \ref{fig:system_interface_overview}, \textit{CoQuest}'s literature discovery feature is presented in the \textit{Paper Graph Visualizer} panel.

When a user clicks
on one of the generated RQ nodes, the \textit{Paper Graph Visualizer} panel reveals itself by visualizing the top-k most relevant papers along with their citation relations retrieved from our citation graph. 
The top-k papers are retrieved using our paper retrieval pipeline, as described in section \ref{sec:paper_retrieval_pipeline}, using the text of the RQ clicked on by the user as the query. 
In the displayed citation graph, each paper node 
represents a paper, and each edge 
represents a citation relation. When the user hovers the cursor over one of the paper nodes, a tooltip will appear with a quick preview of the paper's title information.

The paper nodes in the citation graph are designed to be interactive. When the user clicks 
on one of the paper nodes
, the detailed information of the selected paper will be displayed below. 
The information displayed includes the paper's title, author names, abstract, a TLDR summary provided by Semantic Scholar API\footnote{https://www.semanticscholar.org/product/api}, and a URL linking to the page of the paper on Semantic Scholar. 
Upon the click on the paper node, the \textit{Paper Graph Visualizer} panel will also highlight the selected node and its nearest neighbors (nodes and edges) to indicate which paper(s) have directly cited the selected paper or been cited by the selected paper.

\subsubsection{AI Thoughts: Explaining AI Rationale}
The edges between RQ nodes represent the relation of which RQ the newly generated RQ is based on. They also contain information about the results of the actions undertaken by the agent leading up to the new RQ generated. 
When the user clicks on an edge, the \textit{AI Thoughts} panel appears that displays the results of the agent's action in a narrative format, as shown in the Figure \ref{fig:system_interface_overview}. Detailed implementations of the LLM-based backend for generating RQs and rationales are provided below. 

\begin{figure*}[!t]
    \centering
    \includegraphics[width=\textwidth]{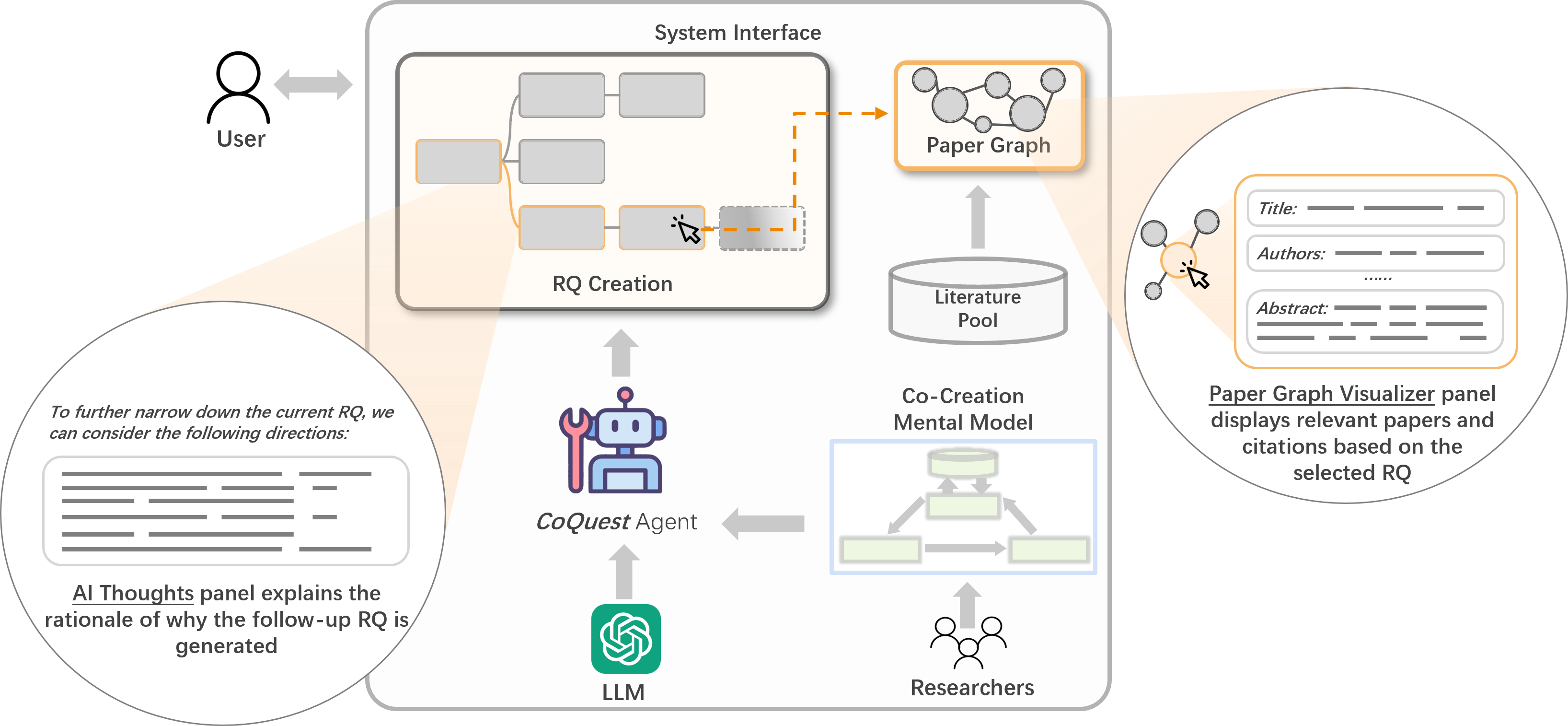}
    \caption{
    \bluehighlight{Illustration of the \textit{CoQuest} framework; The mental model of HCI researchers is used to build the LLM-based agent capable of  accessing and querying a  literature collection to generate research questions (RQs). This agent not only presents the generated RQs to the  users but also provides the rationales behind their generation and literature  grounding through the frontend interface. Examples of prompts used to build the agent can are available in Appendix \ref{apdx:prompt_examples}.}}
    \label{fig:framework}
    \Description{
    This image is a diagram representing a user interface for a research co-creation platform. There are several components illustrated. User: This is likely where the user inputs information or interacts with the system. System Interface: Shows a workflow with several stages, probably indicating how the user can create research questions (RQ Creation) and see a visual representation of related research papers (Paper Graph). AI Thoughts panel: This might be a section where the system provides feedback or suggestions on how to narrow down research questions. CoQuest Agent: Depicts a robot icon, which suggests an AI or machine learning component that assists in the research process. LLM: Stands for Language Learning Model, indicating a machine learning model that processes natural language input. Co-Creation Mental Model: This section includes what appears to be a conceptual framework for how researchers interact with the system. Literature Pool: A database or collection of research literature that the system draws from. Paper Graph Visualizer panel: This panel displays relevant papers and citations based on the selected research question. The diagram is detailed and uses symbols like gears, magnifying glass, and a robot to represent different functions or processes within the platform. The overall design suggests a complex system designed to facilitate collaborative research efforts, possibly through AI-driven insights and a database of academic papers.}
\end{figure*}

\subsection{\textit{CoQuest} Backend and Implementation} 
The backend of the \textit{CoQuest} system comprises an LLM-based agent that automatically generates RQs based on users' input and feedback, by performing reasoning and executing actions that simulate a researcher's mental model in a semi-autonomous manner. 
Two major functions of the \textit{CoQuest} backend include: 1) the RQ generation ability of the LLM agent and 2) the related paper retrieval module that supports literature discovery. 
\bluehighlight{
The framework of how the system's backend connects with the major features are shown in Figure \ref{fig:framework}.
}
\subsubsection{Generating RQs with LLM-based Agent.} 
The \textit{CoQuest} system uses an LLM-based agent to generate creative research questions (RQs) following the ReAct framework \cite{yao2022react}, by adapting the ``\textit{Think}-\textit{Act}-\textit{Observe}'' framework when designing the prompting method. 
First, the "\textit{Think}" step analyzes user input and context to decide an action, resembling human research methods detailed in \ref{sec:mental_model}. During this round, the LLM generates a chain of thought (following our designed prompt) before reaching the conclusion of the action as the next step. 
The actions are executable sub-processes whose results will be used as additional context to help the LLM generate better RQs.
The available actions include: 
1) Search and summarize related works (Literature Discovery); 
2) Hypothesizing use cases (Proposition);
3) Scoping/narrowing down (Refinement);
4) Reflection through comparison with existing works (Evaluation).
Next, during the ``\textit{Act}'' and ``\textit{Observe}'' steps, the execution of actions is achieved in the format of API calls through prompting and can be later parsed and executed through one or multiple pre-implemented Python functions (e.g., retrieve\_papers and summarize\_papers). 
After the next action has been inferred during the ``\textit{Think}'' step, the agent executes the action and appends the results of the action to the context.
Finally, for the agent to generate RQs at each step, we added an additional step of creating RQs at the end of each ``\textit{Observe}'' step. At this step, new RQs are generated based on the provided context and instructions that combine the output from the performed action and predefined prompt.
\bluehighlight{
The detailed usage of prompts in the backend can be found in Appendix \ref{apdx:prompt_examples}.
}

\subsubsection{Related Paper Retrieval.} 
\label{sec:paper_retrieval_pipeline}
In order to help users identify related works more easily, the \textit{CoQuest} system employs a retrieval pipeline to gather existing papers related to the RQs that the users are developing. We curated a literature citation graph from an existing pool of HCI papers, where the nodes represent papers, and the edges represent citation relations.
The \textit{CoQuest} system uses a sentence-based semantic embedding model \cite{reimers2019sentence} to obtain vector representations of each paper in the citation graph. Given a paper's title, metadata, and abstract in text form and a given query (e.g., RQs and user input), the sentence embedding model encodes them into semantic embeddings.
After obtaining the embeddings of papers and the query, the system calculates the similarity between paper candidates and the query. This is done through re-ranking using Maximal Marginal Relevance (MMR) \cite{carbonell1998use}. Then, the system ranks the paper candidates and selects the top-k papers with the highest similarity scores as the final related papers to visualize. More details on the implementation of the retrieval pipeline are discussed in \ref{sec:implementation_detail}.

\subsubsection{System Implementation}
\label{sec:implementation_detail}
The \textit{CoQuest} system is implemented in Typescript as a web application using ReactJS and TailwindCSS for the frontend. The interactive flow editor is implemented using React Flow\footnote{https://github.com/wbkd/react-flow/}. 
The application backend uses Python with FastAPI\footnote{https://github.com/tiangolo/fastapi/}
as the RESTful API server framework. We use AutoGPT\footnote{https://github.com/Significant-Gravitas/Auto-GPT/}
as the foundation of our agent-based LLM implementation\footnote{https://github.com/yiren-liu/coquest}.
We used the \textit{gpt3.5-turbo-16k} model by OpenAI as our LLM engine and \textit{text-embedding-ada-002} model as the sentence embedding model through the cloud service API provided by Microsoft Azure. 
We collected a fixed set of open-access publications through the Semantic Scholar API covering several major HCI conferences (including CHI, CSCW, UIST, Group, IMWUT, IJHCI, and IUI). The final collection of publications includes 2,043 papers.

\section{User Study Evaluating Co-Creation with  \textit{CoQuest}}
To further understand the effect of the \textit{CoQuest} system and how two designs of RQ generation impact users' human-AI co-creation behavior, we conducted a within-subjects user study with 20
\bluehighlight{HCI researchers}
from 8 different institutions by asking the participants to create new research questions using the \textit{CoQuest} system. All participants were graduate students currently enrolled or just graduated with prior experience with research.  
During the study process, we collected participants' behavior and perception (i.e., ratings towards RQs, and ratings towards the \textit{CoQuest} system) data for mixed-method (quantitative and qualitative) analysis. 
All studies were completed remotely online over video calls, where participants were asked to share their screens. Participants were also free to withdraw at any point during the study. Study procedures were approved by the IRB of the researchers' institution. We compensated participants \$20 per hour for the user study. Studies lasted 1-2 hours, including two tasks using two designs (breadth-first and depth-first generations), a survey after each task, and an exit interview. 


\subsection{Within-Subjects User Study - Two Tasks with Assigned Condition: Breadth-First vs. Depth-first}
A within-subjects user study was conducted to understand the difference in users' behavior and perception of the system potentially brought by the two different conditions using the two designs, referred to as breadth-first condition and depth-first condition.
Each participant was asked to complete two tasks: 
During each task, we asked each participant to complete a task designed by researchers to simulate real-life scenarios of research idea formulation. The two task topics used in this study are: \textit{``AR/VR for education and learning''} and 
\textit{``AI and crowdsourcing''}. The two topics were chosen since they cover a wide range of specific domains and provide ample opportunities for users to explore and drill down on related topics.
To account for the effect of the chronological order of both the task topics and conditions on the results, we followed a counterbalanced design \cite{pollatsek1995use} by randomizing the experiment conditions so that all possible orders and combinations were randomly assigned to an equal number of participants. Participants were encouraged to think aloud during the tasks.


\subsection{Data Collection and Analysis} 
We collected both users' perception and behavior data to perform a comprehensive evaluation of the \textit{CoQuest} system.
To understand users' perception of the \textit{CoQuest} system, we gathered two types of perception data from participants: \textit{individual ratings for each generated RQ (RQ ratings)}, which reflects users' perception of their co-creation outcomes; and \textit{overall post-task evaluations of the system (system ratings)}, which reflects users' perception of their co-creation experience. 
Users' behavior data was also collected in the form of system logs and video recordings for later analysis.

\subsubsection{Co-Creation Outcome: Rating AI-Generated RQs During Each Task} The RQ ratings are collected on the fly during each task of the study, where participants were asked to rate at least six RQs of their choice. 
These ratings are intended to capture the immediate perception of users towards the co-creation outcomes (i.e., generated RQs). The during-task rating collection was designed with the intention of nudging users to actively evaluate the RQs during the co-creation process, and also as a way to reflect the accurate user perception in real-time. We adopt Boden’s criteria \cite{boden2004creative} to measure creativity from three different aspects- novelty, value, surprise, and relevance. Participants can give ratings using a 5-point Likert scale slider positioned under each RQ node. 

\subsubsection{Perceived Experience: Survey Scores collected at the End of Each Task, and Exit Interviews} 
To analyze participants' perceived experience using our system, we collected their survey scores upon the completion of each task and conducted exit interviews before the end of each study. 

\textbf{Survey Scores}  
The co-creation experience scores, however, were given by participants at the end of each task through a survey.  The survey contains multiple 5-point Likert scale questions designed to measure their experience from the aspects of: control, creativity, meta-creativity, cognitive load, and trust. We designed 2 Likert-scale questions for each of the 5 aspects to avoid potential bias and ensure a more comprehensive evaluation\footnote{The collected survey ratings have an average Cronbach’s Alpha value of $\alpha = .85$, suggesting good reliability.}. 
The complete list of survey questions can be found in Appendix \ref{apdx:survey_questions}. 

\textbf{Exit Interview}  
After the participant completed the post-task rating survey, we also conducted a ten-minute semi-structured interview to obtain a deeper understanding of the participant's experience with the system. Interview data was analyzed using open-ended coding, by having two researchers review the interview transcripts and frequently discuss with each other \cite{khandkar2009open}. The interview coding highlighted differences in perception between the two conditions explicated by participants.
To measure participants' familiarity with the task topic, we also asked participants how familiar they were with the two topics with three choices: \textit{not familiar}, \textit{kind of familiar}, and \textit{very familiar}.

\subsubsection{Behavior: Think-Aloud Data and System Log.}
We annotated users' behavior (think-aloud transcripts and system usage) to understand how users utilized our system. 

\textbf{Think-Aloud During Co-Creation} Think-aloud data was primarily used for understanding how users generated and interpreted RQs. One researcher first generated a codebook through open coding using videos and transcripts from three randomly selected participants, and then three other researchers independently coded the data of the same three participants, reaching an inter-rater agreement of 0.83 in Krippendorf's alpha. The annotators then discussed and refined the codebook again until they reached full agreement. Then, four researchers proceeded to annotate the remaining 17 participants' behavior data separately. In the final codebook, whether users interacted with the system was annotated and used for quantitative analysis in RQ3 as ``Acted During Wait''. The final codebook also included sense-making behavior (e.g., reasons for (not) waiting, reason for providing certain feedback) as qualitative results.


\textbf{System Log During Co-Creation}
We gathered multiple types of system logs for subsequent analysis of user behavior.
We collected and used the counts of generated RQs and the lengths of user-typed feedback to AI for quantitative analysis of RQ1 and RQ3. User interactions such as clicks on components like RQ nodes and AI Thoughts were used for qualitative analysis along with the think-aloud data.
The text content of user-typed feedback was also used for qualitative analysis in RQ2.

We used different notations throughout this paper to distinguish among the different types of quotes presented in the results. For AI-generated RQs, we use italic (e.g., \textit{AI-generated RQ}); for feedback to the AI, we use double-quotes (e.g., ``feedback''); for interview and think-aloud quotes, we use double-quoted italics (e.g., ``\textit{interview quotes}'').

\begin{figure*}[!t]
    \centering
    \includegraphics[width=\textwidth]{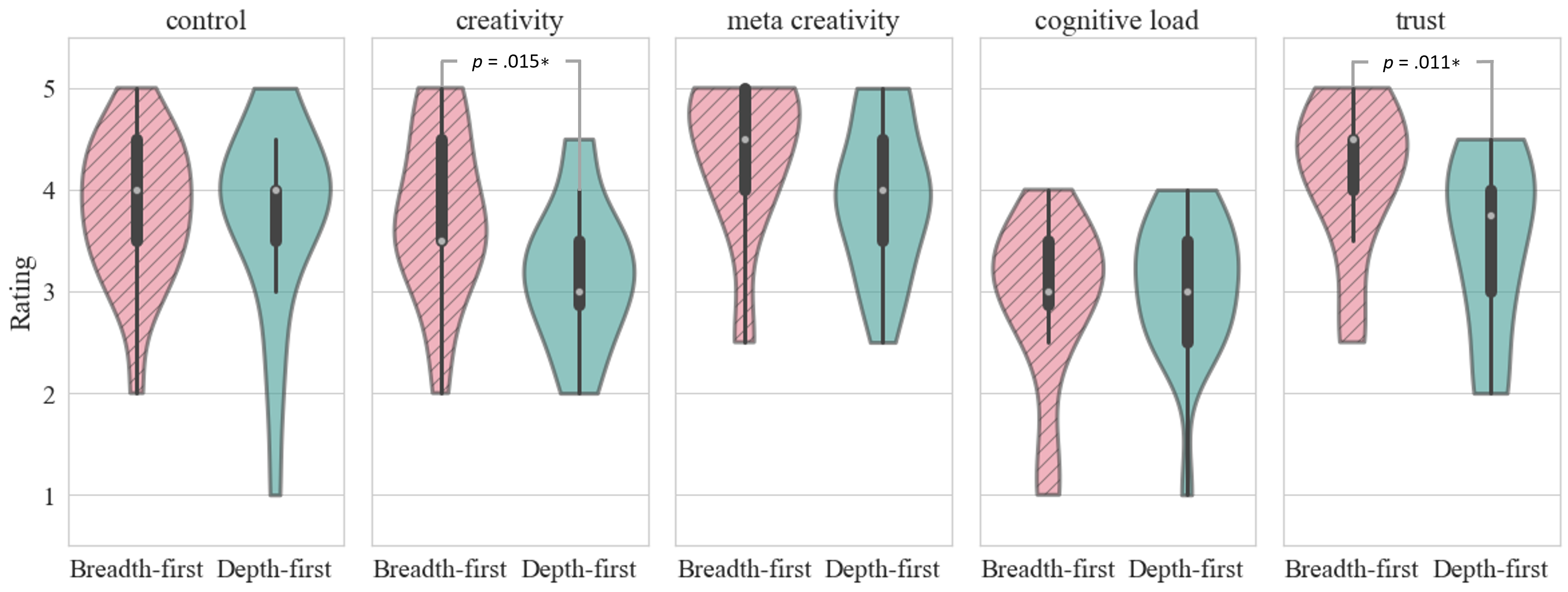}
    \caption{
    Participants rated their perceptions of the system's two designs using a 5-point Likert scale survey. The survey ratings indicated that participants experienced a significantly higher sense of creativity and trust when engaging with the system under the breadth-first condition. 
    }
    \label{fig:survey_boxplot}
    \Description{There are five violinplots (similar to boxplots) that presents users' 5-point Likert scale survey ratings toward five dimensions: control, creativity, meta creativity, cognitive load, trust. Each violinplot has breadth-first and depth-first side by side. For control, the two shapes are very similar with the same median and upper limit. The lower limit is higher for breadth-first than depth-first. For creativity, breadth-first has a higher median and higher upper limit than depth-first, and the lower limit is the same. For meta creativity, the upper limit and lower limit are the same, but breadth-first has a higher median. For cognitive load, the shapes are very similar with the same median, upper-limit, and lower-limit. For trust, breadth-first has a higher median, higher upper-limit, and higher lower-limit than depth-first.}
\end{figure*}

\section{Findings}

\subsection{Perception of Co-Creation Experience and Outcome (RQ1)}

\label{sec:RQ1}




The breadth-first condition allows users to generate multiple RQs in parallel with one interaction, whereas the depth-first condition creates three RQs sequentially, one after another. 
In this section, we analyzed how these two different conditions impact participants' perception towards both co-creation experience and outcomes. 


\subsubsection{Experience: User Perceived Stronger Creativity and Trust Using Breadth-first Condition}
In total, 20 participants created 504 RQs throughout the study, with 276 RQs (M=14.53, SD=6.19) co-created with \textit{CoQuest} system under the breadth-first condition and 228 RQs (M=12, SD=5.31) under the depth-first condition.
To evaluate the overall experience of the co-creation process, we asked our participants to complete surveys upon finishing each task. 
A Mann-Whitney U test\footnote{Power analysis conducted using G*Power\cite{faul2007g}.} of the survey results suggested that users perceived significantly stronger creativity ($U = 288.0$, $p = .015^{*}, d= .68, Power= .83$) using the system under the breadth-first condition (M=3.78, SD=0.82) than the depth-first condition (M=3.18, SD=0.69).
Similarly, the user-perceived trust under the breadth-first condition (M=4.15, SD=0.76) was found significantly higher ($U = 292.0, p = .011^{*}, d= .67, Power= .81$) than the depth-first condition (M=3.5, SD=0.86).
The results of all 5 rating categories are shown in Figure \ref{fig:survey_boxplot}.
\label{stats:survey_higher_creativity}

\begin{figure*}[!ht]
    \centering
    \includegraphics[width=.95\textwidth]{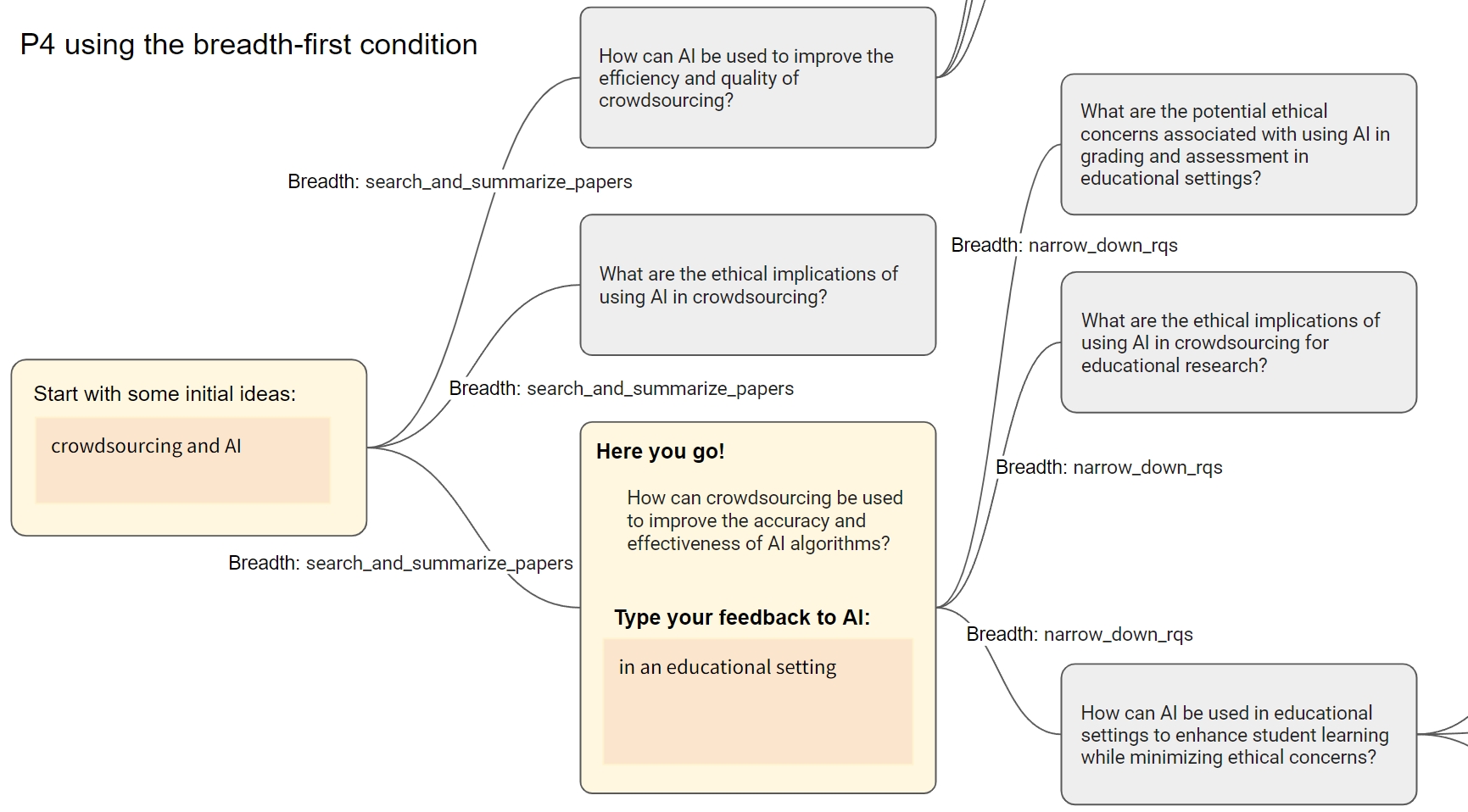}
    \caption{
    Part of P4's RQ flow using the breadth-first condition when exploring the topic of \textit{``AI and crowdsourcing.''} Note that the participant generated from two different RQ nodes in the same iteration, and only one set of generated RQs was presented. The participant provided feedback using keywords such as  ``AI and crowdsourcing'' and ``educational setting'' to help the AI  generate more RQs. The third iteration was not included in this figure.
    }
    \label{fig:P4_breadth}
    \Description{The RQ flow of P4 using breadth-first condition. From the left, there is an initial node with user feedback "crowdsourcing and AI." It leads to three RQ nodes with the same text on each edge "BreadthL search_and_summarize_papers." The generated RQs are: How can AI be used to improve the efficiency and quality of crowdsourcing; What are the ethical implications of using AI in crowdsourcing; How can crowdsourcing be used to improve the accuracy and effectiveness of AI algorithms. The first RQ is connected with three edges, but the three RQ nodes from it are not included in this diagram. The third RQ is expanded with user feedback "in an education setting." This node further leads to three RQ nodes on the right with the same text on edge "Breadth: narrow_down_rqs." The three RQ nodes are: What are the potential ethical concerns associated with using AI in grading and assessment in educational settings; What are the ethical implications of using AI in crowdsourcing for educational research; How can AI be used in educational settings to enhance student learning while minimizing ethical concerns. The third RQ had three more edges connected to it, but generated RQ nodes were not included in this diagram.}
\end{figure*}

During the interview, 12 out of 20 participants (60\%) also mentioned that they preferred the breadth-first condition. 
An example RQ flow generated by the participant (P4) during one of the sessions under the breadth-first condition is shown in Figure \ref{fig:P4_breadth}.
The interview and think-aloud transcripts explained why the breadth-first condition was perceived to create a better experience.

First, the breadth-first condition results are easier to interpret and require less wait time to obtain the same amount of RQs compared with the depth-first condition.
Participants appreciated that the AI was able to list the three generated RQs in parallel ``\textit{all at once}'' (P10), which allowed them to easily ``\textit{compare among the RQs}'' (P10) and ``\textit{explore multiple potential research questions}'' (P4). It is also easier to understand the reason behind the generated RQs in breadth-first condition, as all three RQs shared the same predecessor RQ and rationale.
Although the depth-first condition also generated three RQs using one click, participants had to wait longer to see all three generated RQs than when using the breadth-first condition. 
Therefore, some participants tended to either focus on the first RQ and ignore the other two, or they would start with the last RQ and move to an earlier generated RQ if the latter one was not deemed ideal. 

Second, participants found that the breadth-first condition gave them more control over which direction of RQs they would like to proceed with. 
With the three options listed in parallel, participants were able to choose the RQ that they preferred the most and generate more follow-up RQs based on it. 
The tree-structured design also allowed participants to highlight RQs that were more relevant, and then choose the branch they were more interested in. P14 found the breadth-first condition to be ``\textit{less cognitively demanding},'' and P16 preferred to use the breadth-first condition for ``\textit{brainstorming under one topic}.''
The depth-first condition, on the other hand, may generate RQs in the second and third iterations that were hard to understand how they were related to the first iteration. For example, P6 started an RQ flow with feedback ``crowdsourcing and AI'' with the depth-first condition. While the first question and their feedback did not mention the term \textit{medical diagnosis}, this word appeared after the second iteration but disappeared again after the third iteration. During the interview, they mentioned that the depth-first condition ``\textit{sometimes goes back and forth}'' and it was ``\textit{confusing}'' to understand ``\textit{the logic of why these 3 are parent, child, and grandchild [nodes]}.''

\begin{figure*}[!t]
    \centering
    \includegraphics[width=.95\textwidth]{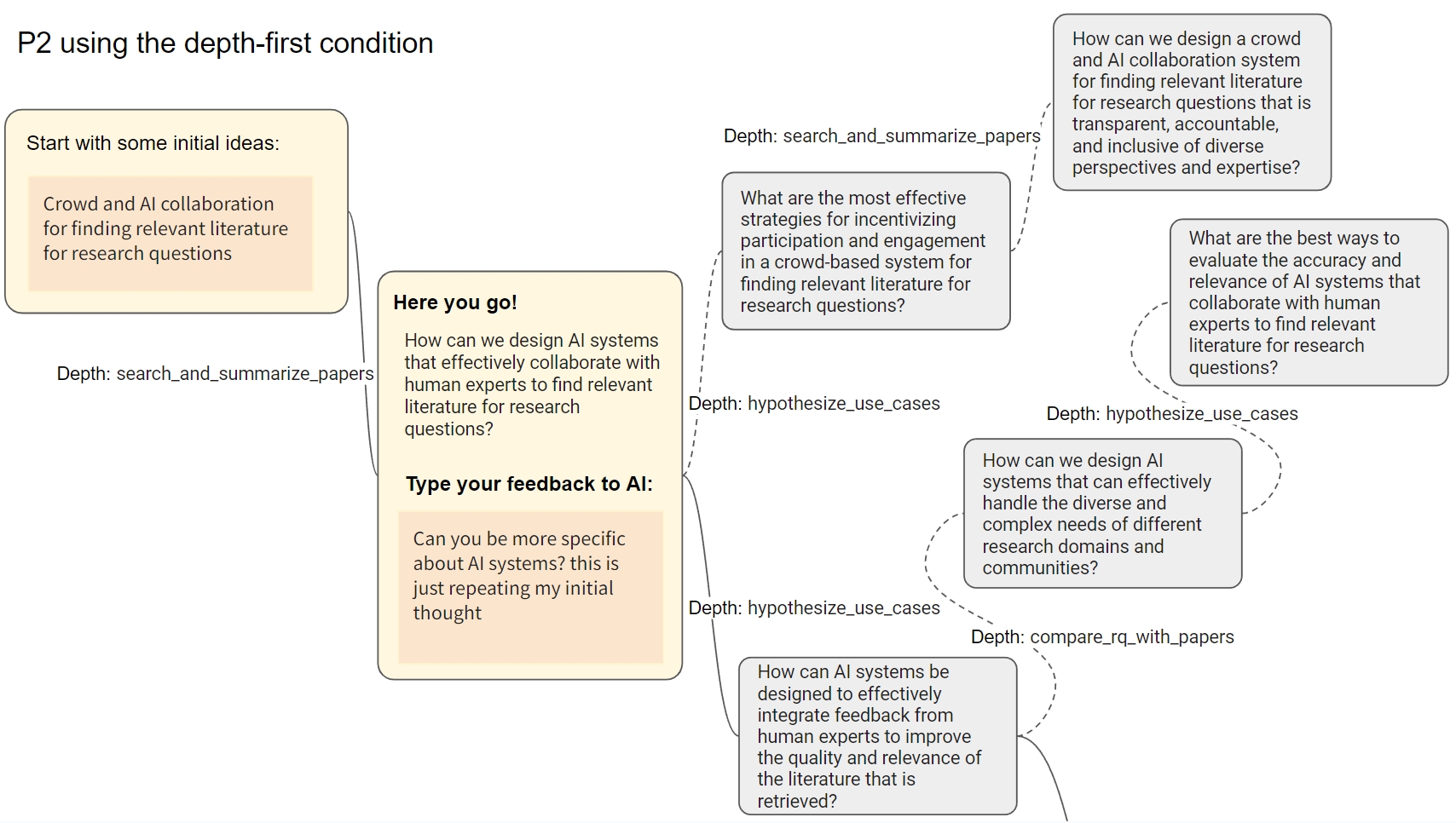}
    \caption{
    Part of P2's \textit{RQ flow} using the depth-first condition when exploring the topic of \textit{``AI and crowdsourcing.''} The dashed lines represent the second and third iterations of RQ generation, which were not based on user feedback. 
    From the first three iterations, the participant chose to continue with the RQ generated in the first iteration, and then provided feedback, asking the AI to be more specific. Subsequently, the AI generated three more RQs, and the participant once again chose to proceed with the first of those  three (subsequent RQs are not included in this figure).
    }
    \label{fig:P2_depth}
    \Description{The RQ flow of P2 using depth-first condition. From the left, there is an initial node with user feedback "Crowd and AI collaboration for finding relevant literature for research questions."  It leads to one RQ node with text "depth: search_and_summarize_papers" on the solid line edge. The RQ is: How can we design AI systems that effectively collaborate with human experts to find relevant literature for research questions. This RQ node is expanded with user feedback "Can you be more specific about AI systems? this is just repeating my initial thought." This RQ node is connected to 2 nodes. The upper node is connected using a dashed edge and text "depth: hypothesize_use_cases." The generated RQ is: What are the most effective strategies for incentivizing participation and engagement in a crowd-based system for finding relevant literature for research questions. From this node, there is another dashed edge connecting to another generated node with text "depth: search_and_summarize_papers" on the edge. The generated RQ is: How can we design a crowd and AI collaboration system for finding relevant literature for research questions that is transparent, accountable, and inclusive of diverse perspectives and expertise. The lower RQ node connected to the first, expanded RQ node is connected with a solid line edge with text "depth: hypothesize_use_cases." The generated RQ is: How can AI systems be designed to effectively integrate feedback from human experts to improve the quality and relevance of the literature that is retrieved. This node is further connected with two edges. The lower edge with a solid line is connected to the outside of the diagram. The upper edge with dashed line has text "depth: compare_rq_with_papers" and connects to the RQ: How can we design AI systems that can effectively handle the diverse and complex needs of different research domains and communities. This node is further connected to another RQ node with a dashed line and text "depth: hypothesize_use_cases" on the edge. The generated RQ is: What are the best ways to evaluate the accuracy and relevance of AI systems that collaborate with human experts to find relevant literature for research questions.}
\end{figure*}

\subsubsection{Outcome: Depth-first Condition Yields RQs with Higher-rated Creativity} 
\label{stats:depth_higher_RQratings}
We measure the outcomes from the co-creation process by asking participants to rate the RQs on the fly during the tasks. To account for the problem of multiple comparisons, we conducted a MANOVA and found a significant difference ($F (4, 209) = 3.79, p = .0057^{**}; Wilk's\ \Lambda = 0.909$) in the user-provided ratings for RQs between the two conditions (i.e., breadth-first and depth-first). 
We conducted Mann-Whitney U tests with these ratings toward generated RQs and found that both the novelty ($U = 2199.5, p = .002^{**}, d=.40, Power=.89$) and surprise ($U = 2387.5, p = 0.017^{*}, d=.32, Power=.75$) of the RQs were rated higher when using the depth-first condition (M=3.78, SD=1.29) compared to the breadth-first condition (M=3.28, SD=1.22).  
In contrast to the post-task survey results that suggested that the breadth-first condition was perceived to be more creative, the RQ ratings suggested that the generated RQs using the depth-first condition were more innovative instead. Figure \ref{fig:P2_depth} shows an example of participants using the depth-first condition.

During tasks and interviews, participants explained why they found the outcomes from the depth-first condition to be more innovative. 
First, participants found that the depth-first condition tended to generate surprising RQs with ideas that were not present previously or included in their feedback to AI.
Since the second and third RQs were generated based on the first RQ without user feedback, the AI sometimes may add unexpected keywords to the RQs. While a few participants found these unexpected additions to be ``\textit{surprising in a negative manner}'' (P3) or ``\textit{distracting}'' (P12), more participants described it in a positive way such as ``\textit{insightful}'' (P18) or ``\textit{impressive}'' (P8) and found it interesting to see the AI ``\textit{thinking when generating the questions} (P11).'' This would be particularly useful if the participant was less familiar with a certain topic, as it would be more inspiring. 

Second, the depth-first condition allows participants to dive deep into one chosen RQ and improve its quality. While it requires ``\textit{more patience}'' (P14) to wait for and read through all three generated RQs from one interaction using depth-first condition, participants found it easier to go  ``\textit{deeper}'' and form an RQ that is ``\textit{more specific}'' (P4). The depth-first RQs were also found to be more ``\textit{creative and unique}'' (P13). 
P17 also expressed a positive opinion about having the freedom to choose and generate follow-up RQs from any AI-generated RQs they would like to work on.
In fact, participants could also utilize the AI generation wait time to better interpret existing RQs and generate more valuable RQs. 


\begin{figure*}[!t]
    \centering
    \includegraphics[width=.95\textwidth]{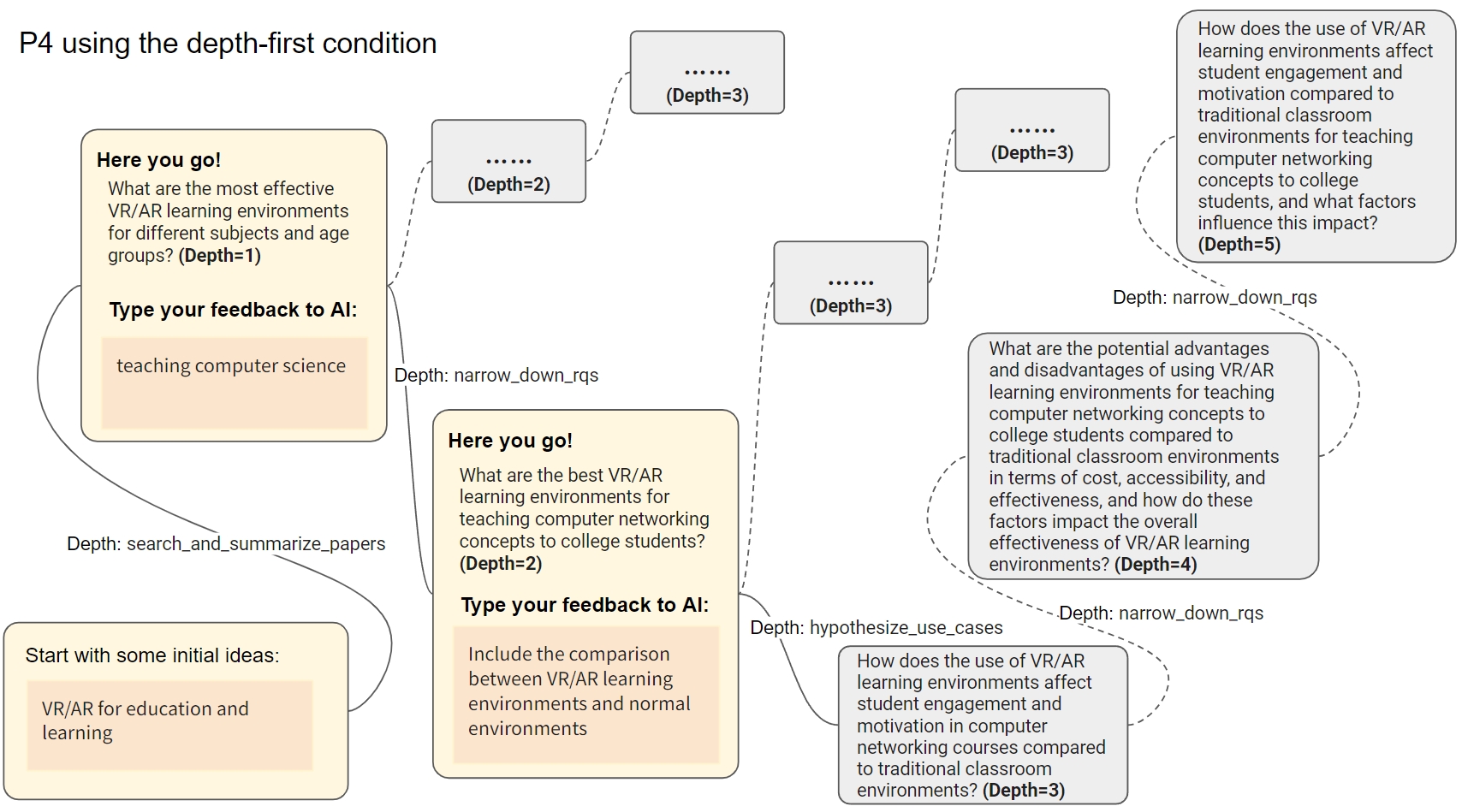}
    \caption{
    \bluehighlight{
        Part of P4’s \textit{RQ flow} involved using the depth-first condition when exploring the topic of ``\textit{AR/VR for education and learning}.'' The dashed lines represent the second and third iterations of RQ generation, which were not based on the user's feedback. The participant provided feedback to the RQs generated at depth=1 and depth=2. Based on the RQ at depth=2,  three more RQs were generated. P4's perceived novelty, surprise, value, and relevance initially scored at (3,4,3,5) during the first iteration (at depth=1), and these scores increased to (4,5,4,5) as the depth increased.
        }
    }
    \label{fig:rqExampleDepthVar}
    \Description{The RQ flow of P4 using depth-first condition. From the left, there is an initial node with user feedback "VR/AR for education and learning."  It leads to one RQ node with text "depth: search_and_summarize_papers" on the solid line edge. The RQ is: What are the most effective VR/AR learning environments for different subjects and age groups? This RQ node is denoted as depth=1. It is expanded with user feedback "teaching computer science." This RQ node is connected to 2 nodes. The upper node, denoted as depth=2, is connected to using a dashed edge, and further connected to another node, denoted as depth=3, using a dashed edge. Both edges did not have text and the RQs in these nodes were omitted. The lower node was connected with a solid line edge with text "Depth: narrow_down_rqs." The generated RQ is: What are the best VR/AR learning environments for teaching computer networking concepts to college students? This node is denoted as depth=2. This RQ node is connected to 2 nodes. The upper node, denoted as depth=3, is connected to using a dashed edge, and further connected to another node, denoted as depth=4, using a dashed edge. Again, both edges did not have text and the RQs in these nodes were omitted. The lower node was connected with a solid line edge with text "Depth: hypothesize_use_cases." The generated RQ is: How does the use of VR/AR learning environments affect student engagement and motivation in computer networking courses compared to traditional classroom environments? This node is denoted as depth=3. From this node, there is a dashed edge connecting to a generated node with text "Depth: narrow_down_rqs" on the edge. The generated RQ is: What are the potential advantages and disadvantages of using VR/AR learning environments for teaching computer networking concepts to college students compared to traditional classroom environments in terms of cost, accessibility, and effectiveness, and how do these factors impact the overall effectiveness of VR/AR learning environments? This node is denoted as depth=4. This node is further connected to another node using dashed lines with text "Depth: narrow_down_rqs" on the edge. The generated RQ is: How does the use of VR/AR learning environments affect student engagement and motivation compared to traditional classroom environments for teaching computer networking concepts to college students, and what factors influence this impact? This node is denoted as depth=5.}
\end{figure*}

\bluehighlight{
To further understand how users' perception towards RQs varied throughout the co-creation process, we performed a temporal analysis over the depth of RQ flows. 
By computing Spearman's correlation coefficient, we found significant positive correlations between the depth of RQs and their corresponding ratings of novelty ($r(514) = .41$, $p < .001^{***}$), value ($r(514) = .22$, $p = .011^{*}$) and surprise ($r(514) = .24$, $p = .006^{**}$). We also found that RQs with depths of 9 or higher seemed to exhibit a drop in ratings. This might be due to the repetitiveness of RQs that was caused by the narrowing-down of topic and thus fixed literature space.
To further illustrate this narrowing-down process of RQs, a sample flow of RQs generated under depth-first condition by one of the participants, P4, is shown in Figure \ref{fig:rqExampleDepthVar}. The participant initially expressed confusion towards the first RQ generated as it was perceived as ``kind of vague'' and ``more like a literature review question instead of research question.'' After instructing the system to provide more details, P4 read and commented on the follow-up RQ generated as ``an OK research question… But I wish it could be more specific on what (factor) makes VR/AR good for learning environment…'' Then, they generated three more follow-up RQs and perceived them to be more specific and surprising ``I’m a bit surprised that it was able to pick up the concept of accessibility.'' However, they noted that making sense of the connections between later RQs was more challenging.
}

\textbf{Summary (RQ1):} 
According to the post-session survey results, the breadth-first condition offered users a better co-creation experience, with higher perceptions of creativity and trust. 
In contrast, RQ rating data showed that the depth-first approach led to the generation of RQs deemed more creative by participants, with a unique depth and unexpected novelty. 
This insight indicates that while the breadth-first condition enhances user engagement, the depth-first condition stimulates deeper and more novel outcomes, leading to distinct advantages in different aspects of the co-creation process.
\bluehighlight{
Positive correlation between users' ratings towards RQs and depth were also observed, indicating improved user perception towards co-creation outcomes as exploration unfolded. 
}

\subsection{Users' Co-creation Behavior with LLM-based Agent (RQ2)}
To explore users' co-creation behavior with AI, we examined how they interacted with the \textit{CoQuest} system through think-aloud comments, feedback to AI, and activities performed while they were waiting for AI to generate the results. 
In this section, we explain the findings to shed light on how participants used the \textit{CoQuest} system under the two conditions.

\subsubsection{Depth-first Condition Stimulated More User Interactions During Wait Time}
\label{sec:wait_time}
The \textit{CoQuest} system required users to wait for approximately 30 seconds for the LLM inference to finish after each time the generation was triggered. During this wait time, users can utilize other system features in parallel to the generation, such as interpreting other RQs, generating additional RQs, viewing \textit{AI Thoughts}, and exploring the paper graph.
By examining the behavior data of participants, we found that most participants utilized the wait time during the generation of RQs to perform other activities. 
In total, 12 out of 20 participants utilized the wait time to interpret and evaluate other existing RQs on the RQ flow editor. Among them, 7 participants created new RQs while waiting for prior RQs generation to be finished. 
We also found that more participants explored other RQs during wait time under the depth-first condition (N=12) than the breadth-first condition (N=6), with a two-proportion z-test indicating significance ($z = -2.01$, $p = 0.045^{*}$).

Participants were observed to switch among different threads of RQs during the wait time of generation. 
P16, for example, utilized wait time under depth-first conditions to start generating another thread of RQs. 
The participant further explained during the interview that if the RQs were generated quicker, ``\textit{probably would just end up exploring one at a time. But because it was taking a while, I was like, oh, I can write another one}.'' The participant found the generation wait time beneficial for brainstorming new ideas, as it offered a good opportunity for exploring different directions in parallel. 
In addition, we also found a positive correlation, shown by Pearson's correlation test ($r(392) = 0.18, p < 0.001^{***}$), between whether a user made use of the RQ generation wait time and the length of the user's feedback to AI. 
The test result indicated that users utilized the RQ generation wait time to contemplate more detailed feedback for AI, suggesting higher engagement in human-AI co-creation.




During the wait time for generation, participants were also found to use the ``AI thoughts'' feature to understand the relationship between different RQs and the rationales behind generated RQs. 
Though it was confusing for some participants, most participants were observed to have read the ``AI thoughts,'' especially while waiting for all three RQs to be generated under the depth-first condition (14 out of 20 users clicked and viewed the ``AI thoughts'' more than 10 times).
More specifically, in the depth-first condition, each click generates three different ``AI thoughts'' compared to breadth-first condition, where each click only generates one ``AI thoughts.'' In addition to understanding the logical relation and reasoning behind RQs, participants also used ``AI thoughts'' for other co-creation purposes that were shared between the two conditions. For instance, users have found AI thoughts to be helpful both in terms of understanding the rationale of how RQs are generated, and also providing additional ideas for users to elaborate on. 
Some participants used the AI thoughts for sense-making of the generated RQs. When they recognized the AI thoughts were connected with the paper graph, they expressed that they trusted the system more because it seems like the AI thoughts are ``\textit{from actual papers}'' that ``\textit{are peer-reviewed}'' rather than  ``\textit{coming from large language model, which is not necessarily [factually] true}'' (P10). Among our different types of ``AI thoughts,'' one participant who could tell such differences (P16) said they liked the ``\textit{summary of existing work}'' explanation type because it seems like this type ``\textit{builds on existing works}'' and makes it more trustworthy.
Some other participants also used the terms and keywords that appeared in ``AI thoughts'' to help develop their feedback to further guide RQ generation. For example, P10 wrote the feedback to AI ``Explore feature selection for non-experts'' by taking inspiration from the AI's thoughts about a summary of works related to feature selection techniques for data mining.


\bluehighlight{
To understand how users’ actions during wait time varied temporally during the exploration process, we performed Mann-Whitney U tests to examine the difference between users' wait time actions under breadth- and depth-first conditions within each stage, i.e. earlier (depth<=3) and later (depth>3) stages. The stage of exploration is determined by computing the mean depth across all RQ nodes ($M=3.38$).
Significant behavioral differences were observed between depth-first and breadth-first design conditions. The earlier stage of interaction revealed notable distinctions: users in the breadth-first condition engaged more frequently in 'Check paper graph' ($U = 4414$, $P = 0.009^{**}$) and 'Generate new RQs' ($U = 4483$, $P = 0.011^{*}$) actions compared to those in the depth-first condition. This suggests that the breadth-first approach, which presents multiple research questions simultaneously, encourages more active exploration and engagement early in the interaction.
However, these differences diminished in the later stage, indicating a convergence in user behavior as interaction progresses. This finding highlights the impact of initial information presentation on user actions, particularly in systems like \textit{CoQuest} where engagement patterns can influence the user's exploratory experience.
}

\subsubsection{Users Have Different Strategies for Providing Feedback to AI}
\label{sec:different_strategy_feedback}
The \textit{CoQuest} system allows users to type in textual feedback to AI under each RQ before triggering the generation of new RQs. 
Overall, 20 participants wrote 119 pieces of feedback in total using both conditions (M=7.44, SD=5.25) during the co-creation process. 
To find whether the participants had different ways of communicating with the AI using text feedback, we first analyzed the difference among users' input lengths (word counts) by fitting a mixed-effects model that considers both the random effect from different users and the potential fixed effect from the two conditions. 
Details of the model and results can be found in Appendix \ref{apdx:mixed_effect_model}. 
A likelihood ratio test was then conducted to validate the significance of the random effect brought by users ($\chi^2(1) = 9.189, p = 0.002^{**}$). The result of the likelihood ratio test indicated a statistically significant difference in feedback lengths among different users regardless of the condition.
However, we found no significant association between the system condition and users' feedback lengths. 


To obtain a deeper understanding of how participants employ different strategies in providing feedback to AI, we further reviewed and coded participants' feedback to AI during the co-creation process and observed that participants tend to provide feedback to the AI in different ways. More specifically, we found three main themes.

The most common theme of giving AI feedback is to provide keywords (N=59), which is shorter in length. For feedback under this theme, 61.0 \% (N=36) were used to start a new node using the keywords that the participant would like the AI to start with. From a generated RQ, participants may also provide new keywords and instruct the AI to include these keywords in the next iteration. Search and summarize was frequently used by AI to generate RQs based on the new keywords. For example, P4 started a node using the breadth-first condition with ``AI and crowdsourcing,'' as shown in Figure \ref{fig:P4_breadth}. The participant then continued with the generated RQ \textit{What are the ethical implications of using AI in crowdsourcing?} and wrote ``in an educational setting.'' The AI then generated three new RQs with the new keyword.

The second theme is asking AI-specific questions (N=40). These questions are usually longer and in full sentences that either ask the AI to explain terms used in a generated RQ, to lead the AI toward a new direction, or to ask the AI to review more related literature. For example, based on the RQ \textit{What is the impact of medical simulations on medical student learning and skill development?}, P17 wrote, ``It is a good question to research into, can you give me some information that is common among the papers.'' The participant was able to proceed with one of the three generated RQs.

A third theme is asking the AI to be more specific on an RQ (N=55). Feedback under this theme usually starts with the phrase ``\textit{be more specific}.'' In 45 out of the 55 cases, participants would also include the direction that they would like the AI to continue on. In response, the AI usually narrows down RQs or performs search and summarization. For example, after seeing the RQ \textit{How can we design AI systems that effectively collaborate with human experts to find relevant literature for research questions?} using depth-first condition, P2 wrote, ``Can you be more specific about AI systems? this is just repeating my initial thought.'' The AI then generated three more RQs, as shown in Figure \ref{fig:P2_depth}.

\textbf{Summary (RQ2):} 
Participants were found to have employed different strategies when co-creating research questions (RQs) with AI.
Participants mainly provided feedback by listing keywords, posing additional questions, or requesting specificity. While waiting for the AI's response, users often explored other RQs, showing greater engagement in conditions with longer wait times. The \textit{AI Thoughts} panel, which offers insights into the AI's reasoning, was found beneficial in enhancing trust and inspiring feedback, especially when grounded in peer-reviewed literature.
\bluehighlight{
Moreover, our analysis revealed that in the early stages of interaction, users in the breadth-first condition engaged in more actions during wait time, such as checking paper graphs and generating new research questions, indicating that initial information presentation in a breadth-first manner notably influenced users' early exploration behavior.
}

\subsection{Association between Users' Persona/Behavior and Their Perceptions (RQ3)}


\begin{table*}[!h]
    \centering
    \caption{
    Regression Results with RQ ratings as dependent variables and users' behavior data as predictors. Each column in the table represents one regression performed with the corresponding rating item as the dependent variable.
    }
    \label{Tab:reg_RQrating_behavior}
    \begin{tabularx}{\textwidth}{lRRRR}
        \toprule
        & \multicolumn{4}{c}{\textbf{Dependent Variables --- Outcomes (RQ Ratings)}} \\
        \cmidrule(lr){2-5}
        \textbf{Predictors} & Novelty\phantom{\textbf{*}} & Value\phantom{\textbf{*}} & Surprise\phantom{\textbf{*}} & Relevance\phantom{\textbf{**}} \\
        & $\beta$\ (S.E.)\phantom{\textbf{*}} & $\beta$\ (S.E.)\phantom{\textbf{*}} & $\beta$\ (S.E.)\phantom{\textbf{*}} & $\beta$\ (S.E.)\phantom{\textbf{**}} \\
        \midrule
        Condition (Depth-First=1) & \textbf{1.53 (.63)*} & \textbf{1.54 (.64)*} & \textbf{1.52 (.60)*} & \textbf{1.70 (.68)*}\phantom{\textbf{*}} \\
        Feedback Length & \textbf{.15 (.06)*} & \textbf{.17 (.06)*} & \textbf{.15 (.06)*} & \textbf{.18 (.06)**} \\
        Total \# of RQs Created & -.01 (.03)\phantom{\textbf{*}} & -.03 (.03)\phantom{\textbf{*}} & -.02 (.03)\phantom{\textbf{*}} & -.04 (.04)\phantom{\textbf{**}} \\
        Acted During Wait & -.06 (.07)\phantom{\textbf{*}} & -.07 (.07)\phantom{\textbf{*}} & -.06 (.06)\phantom{\textbf{*}} & -.08 (.07)\phantom{\textbf{**}} \\ 
        Familiarity & .19 (.14)\phantom{\textbf{*}} & .18 (.14)\phantom{\textbf{*}} & .20 (.13)\phantom{\textbf{*}} & .22 (.15)\phantom{\textbf{**}} \\
        \bottomrule
        \multicolumn{5}{l}{\footnotesize *: p$<$0.05, **: p$<$0.01, ***: p$<$0.001}
    \end{tabularx}
\end{table*}

\begin{table*}[!h]
    \centering
    \caption{Regression Results with survey ratings as dependent variables and users' behavior data as predictors.}
    \label{tab:reg_survey_behavior}
    \begin{tabularx}{\textwidth}{lRRRRR}
        \toprule
        & \multicolumn{5}{c}{\textbf{Dependent Variables --- Experience (Survey Scores)}} \\
        \cmidrule(lr){2-6}
        \textbf{Predictors} & Control\phantom{\textbf{**}} & Creativity & Meta Creativity & Cognitive Load & Trust \\
        & $\beta$\ (S.E.)\phantom{\textbf{**}} & $\beta$\ (S.E.) & $\beta$\ (S.E.)\phantom{\textbf{**}} & $\beta$\ (S.E.)\phantom{\textbf{***}} & $\beta$\ (S.E.)\\
        \midrule
        Cond. (Depth-First=1) & .34 (.71)\phantom{\textbf{**}} & -.41 (.46)\phantom{\textbf{*}} & \textbf{-.74 (.23)**} & \textbf{.82 (.36)*}\phantom{\textbf{**}} & -.58 (.88)\phantom{\textbf{*}} \\
        Feedback Length & -.03 (.04)\phantom{\textbf{**}} & .02 (.05)\phantom{\textbf{*}} & -.01 (.03)\phantom{\textbf{**}} & \textbf{.07 (.01)***} & .03 (.04)\phantom{\textbf{*}} \\
        Total \# of RQs Created & \textbf{-.07 (.03)*}\phantom{\textbf{*}} & -.01 (.01)\phantom{\textbf{*}} & \textbf{.03 (.01)*}\phantom{\textbf{*}} & -.01 (.01)\phantom{\textbf{*}}\phantom{\textbf{**}} & .02 (.05)\phantom{\textbf{*}} \\
        Acted During Wait & \textbf{.10 (.04)**} & -.06 (.07)\phantom{\textbf{*}} & -.03 (.04)\phantom{\textbf{**}} & .06 (.05)\phantom{\textbf{*}}\phantom{\textbf{**}} & -.03 (.04)\phantom{\textbf{*}} \\
        Familiarity & \textbf{-.22 (.08)**} & .04 (.09)\phantom{\textbf{*}} & -.09 (.05)\phantom{\textbf{**}} & \textbf{-.08 (.04)*}\phantom{\textbf{**}} & .07 (.08)\phantom{\textbf{*}} \\
        \bottomrule
        \multicolumn{6}{l}{\footnotesize *: p$<$0.05, **: p$<$0.01, ***: p$<$0.001}
    \end{tabularx}
\end{table*}

To better understand the factors influencing participants' perceived experiences and outcomes, we performed two linear regression analyses, aiming to associate user perceptions with their behaviors and backgrounds.
The two sets of regressions aim to yield insights about users' perceptions from two perspectives: 1) the first set of regressions adopts participants' ratings for RQs (i.e., novelty, value, surprise, and relevance) as dependent variables; 2) another set of regressions participants' post-session survey scores for the system (i.e., control, creativity, meta creativity, and cognitive load) as the dependent variables. 

We chose factors that could represent users' behavior as predictors based on our previous findings\footnote{
To examine the potential issue of multicollinearity, we calculated the Variance Inflation Factor (VIF) for each predictor. All predictors yielded VIFs lower than the common threshold of 5 \cite{james2013introduction}, indicating no substantial collinearity problem.
}. 
We first considered the conditions of the system (breadth-first or depth-first) experienced by the user, as indicated by the results in \ref{sec:RQ1} that the two conditions might have a difference between their impact over the user's perception of co-creation experience and outcomes. 
We constructed the predictor as a categorical variable, with value 0 standing for breadth-first condition and value 1 standing for depth-first condition. 
We also considered factors related to user engagement, including the length of user feedback to AI, which was found to vary across different users as mentioned in \ref{sec:different_strategy_feedback}, and the count of total RQ nodes created. 
Additionally, we included whether users performed any actions while waiting for results from an already triggered generation as a potential factor, as discussed in \ref{sec:wait_time}.
Participants' familiarity with task topics was also considered as a predictor, measured through the familiarity ratings collected through the post-session surveys. 
The familiarity variable was constructed as an ordinal variable taking three possible integer values from 0 (not familiar) to 2 (very familiar).

\textbf{Positive perceptions of generated RQs associated with longer feedback to AI.}
As shown by the results in Table \ref{Tab:reg_RQrating_behavior}, a positive association has been identified between all four dimensions of user-perceived RQ ratings and the average lengths of users' input feedback lengths. The finding suggests that increased user engagement in the human-AI co-creation process through providing textual feedback might improve the perceived outcome quality.
In addition, we found that the RQ ratings are positively associated with whether a user used the system under the depth-first condition. Namely, users perceived the RQs generated by AI under the depth-first condition to be of better quality than the breadth-first condition. This also corresponds to our findings in \ref{stats:depth_higher_RQratings}. 
From Table \ref{Tab:reg_RQrating_behavior}, it can also be seen that participants' feedback length is positively associated with perceived cognitive load. 
The results indicate that, although providing longer feedback required users to invest more thoughts, it also improved the co-creation outcomes. 
The finding also aligns with our earlier qualitative results in \ref{sec:RQ1}. The interview result further reveals that providing feedback promoted their reflection on themselves and trying to understand how AI works, although tiring, simultaneously helped them improve and manage their own creative thinking processes. 
More specially, some participants were observed during think-aloud performing ``reverse engineering'' to interpret how AI works and even try to replicate the generation process in a new thread. 

\textbf{Factors associated with users' perception of co-creation experience.}
Table \ref{tab:reg_survey_behavior} shows the regression results with participants' survey rating responses as dependent variables and user-specific factors as predictors. 
We found a statistically significant positive association between whether a participant has utilized the RQ generation wait time to perform other activities and the participant's perceived control for the system. This indicates that users tend to perceive higher control of the system when better utilizing the generation wait time beyond merely waiting for the generated results. 
There was also a negative association between the total number of nodes generated by participants and their perceived control, which indicates that as users generate more RQs during each session, they perceive to have weaker control of the system. 

Participants' familiarity with the task topics was also found to be negatively associated with their perceived control. This was also reflected by participants during their think-aloud process, as users who were more familiar with the task topic had stronger expectations for the generated RQs: 
\userquote{... in educational research, we don't say crowdsourcing anymore. We say learner sourcing ...
} (P4). 
We also noted that more experienced researchers would explicitly specify their needs for the agent to generate RQs in directions using terms related to research methodology. For example, one feedback P17 wrote to AI was ``What the metrics for effectiveness and engagements.''
Several participants reflected that they were having a rough initial thought in their mind when being assigned a topic that is familiar to them, and what AI does is only help them externalize the thoughts. P4, for example, found several generated RQs to be ``surprising in a negative manner'' as he would evaluate whether an RQ ``\textit{is something interesting, something novel, and ... relates to some of the prior literature}.''

\textbf{Summary (RQ3):} 
Factors related to participants' system usage were found to be associated with users' perceptions of co-creation experiences and outcomes.
The depth-first condition was also associated with better-perceived outcome quality in AI-generated RQs than the breadth-first condition. 
Users' familiarity with task topics, interestingly, led to a reduced sense of control, indicating users with different levels of expertise may have specific expectations unmet by the AI system.
Additionally, we found that users providing longer feedback to AI, although introducing a larger cognitive load, led to outcomes with better quality.

\section{Discussion}
In this section, we explore how the insights from our user study align with established theories and previous research. Furthermore, we offer design implications for future co-creation systems using Large Language Models.

\begin{figure*}[!h]
    \centering
    \includegraphics[width=0.9\textwidth]{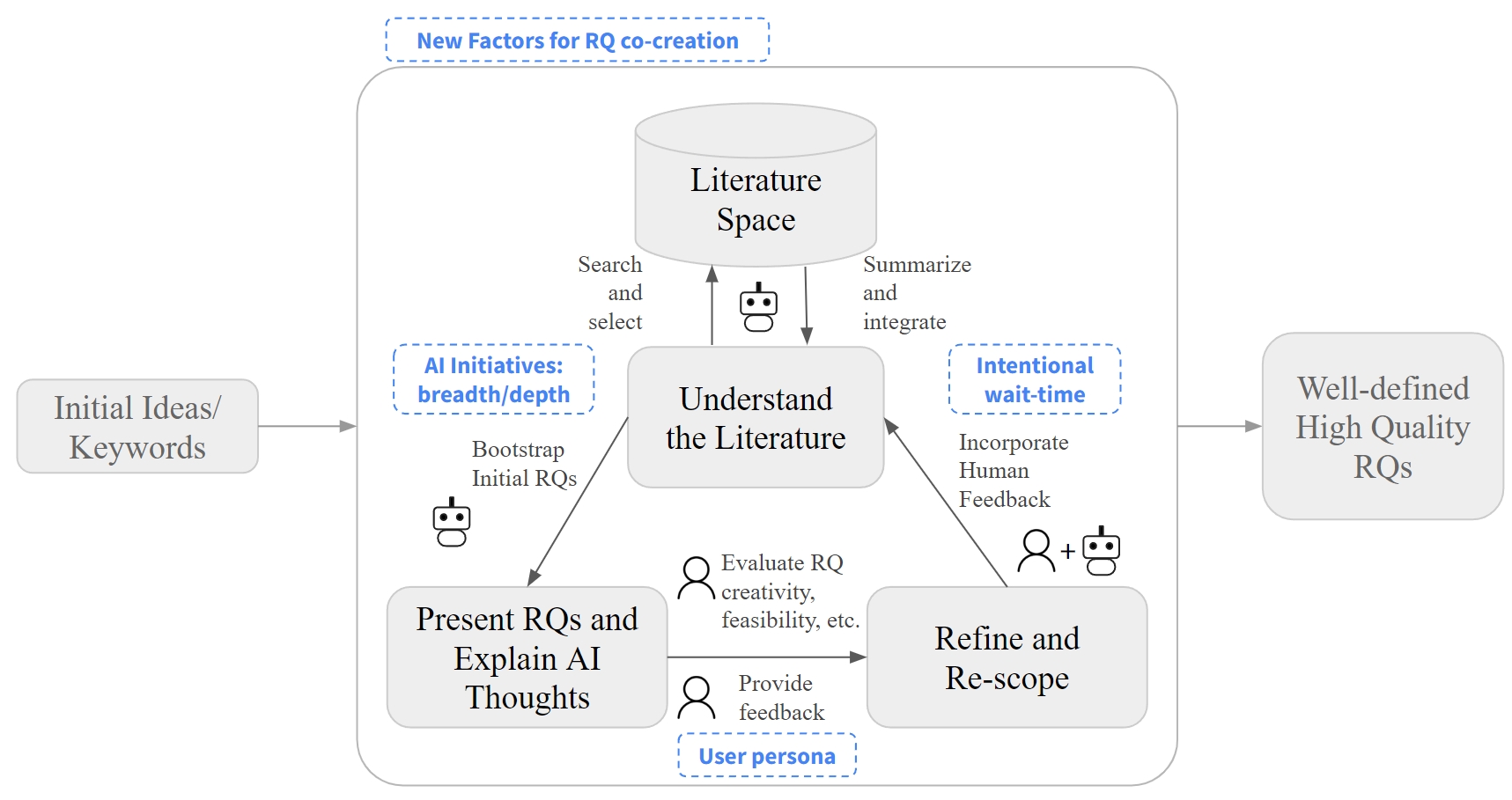}
    \caption{
    Updating the mental model of human-AI RQ co-creation with additional factors identified through the experimental  study.
    }
    \label{fig:mental_model_plus}
    \Description{This is the updated human's mental model of RQ co-creation with the LLM-based agent. The following description are the same as the initial model in Figure. There are four components inside this model: Present RQs and Explain AI Thoughts (lower left, rectangular), Understand the Literature (center, rectangular), Literature Space (top, with cylinder shape), and Refine and Re-scope (lower right, rectangular). From Present RQs and Explain AI Thoughts to Refine and Re-scope, there are two user-driven activities: Evaluate RQ creativity, feasibility, etc., and Provide feedback. From Refine and Re-scope to Understand the Literature, there is one user- and AI-driven activity: Incorporate Human Feedback. From Understand the Literature to Present RQs and Explain AI Thoughts, there is one AI-driven activity: Bootstrap Initial RQs. From Understand the Literature to Literature Space, there is one AI-driven activity: Search and select. From Literature space to Understand the Literature, there is one AI-driven activity: Summarize and Integrate. The model has "Initial idea / keywords" as input and "Well-defined High Quality RQs" as output. The following descriptions are the differences from the initial model. There are three additional blue texts highlighting the additional factors added to the model. From Present RQs and Explain AI Thoughts to Refine and Re-scope, there is blue text: User persona. From Refine and Re-scope to Understand the Literature, there is blue text: Intentional wait-time. From Understand the Literature to Present RQs and Explain AI Thoughts, there is a blue text: AI Initiatives: breadth/depth.}
\end{figure*}

\subsection{Enriching Our Proposed Mental Model for Human-AI Co-Creation of Novel Research Questions}
Existing studies \cite{foster2004nonlinear, palmer2009scholarly} have mostly been focused on the research lifecycle as a whole when discussing the models of the research process. 
Our user study findings shed light on additional factors to consider when designing future human-AI co-creation systems for research question generation, as shown in Figure \ref{fig:mental_model_plus}.


\subsubsection{Wait Time in AI-based Co-Creation System as an Opportunity to Promote Creativity}
Our observation of users' behavior patterns when using the \textit{CoQuest} system sheds light on the possibility of utilizing the AI system's processing wait time as a design opportunity for users' exploration in the context of human-AI co-creation systems.
The RQ3 findings revealed that when users utilized wait time to perform other activities, it improved their perceived control of the co-creation experience. This is contradictory to the common belief that the response delay in AI-based systems only leads to a negative impact on users' experience. 
\bluehighlight{In our RQ2 findings, two participants (P10 and P16) mentioned that \textit{AI Thoughts} panel improved their trust for the system. However, this effect may not have been significant in the regression results of Table \ref{tab:reg_survey_behavior} due to the aggregation of various actions under the ``acted during wait'' category. We recognize that the impact of the AI thoughts panel on trust might have been diluted when combined with other actions in the aggregated data, thus statistically insignificant. Future work should separate these actions to measure their individual impacts on trust more precisely. This will allow a clearer understanding of how specific features, such as the AI thoughts panel, contribute to enhancing user trust.}
Our RQ2 results also showed that users utilized the time waiting to jump across different threads of RQs and perform exploration of ideas in parallel, or to articulate more detailed feedback for AI.
The utilization of wait time was found to be prominent especially under the depth-first condition, which could be one of the reasons leading to its higher perceived co-creation outcome quality.

Most recent studies on LLM optimization focus on reducing the inference time and speeding the generation wait time needed for general purposes \cite{liu2023deja, leviathan2023fast, dettmers2023case}. 
Our findings, however, provided a unique perspective that in the context of LLM-based co-creation systems, the generation wait time can be exploited, or even purposely introduced, to promote users' creative activities. 
This can be achieved using different design techniques, such as tree-based visualization that highlights concurrency, or incorporating interactive nudges that guide users towards reflection, ideation, or brainstorming during these pauses. Such techniques can not only enhance user engagement but also maximize the cognitive benefits derived from the wait time breaks, potentially leading to more innovative and diverse co-creation outcomes.

\subsubsection{Aligning Users' Perception towards Experience and Outcomes of Generated RQs for Improved Creativity}

In addition to prior studies' understanding of mix-initiative system design \cite{rezwana2022designing, weisz2023toward}, our study provides empirical findings to support the need to consider the degree of initiative taken by AI as a design option in further human-AI co-creation system design.
Our RQ1 findings indicate that the degree of initiative taken by AI during co-creation could affect users' perception of both co-creation experience and outcomes: If AI takes less initiative and gives users more freedom to choose from various generation outputs, it improves the users' co-creation experience. Contrarily, if AI drives deeper thoughts by taking more initiative, it leads to co-creation outcomes with higher quality and creativity. Based on our study results, this can be implied through lower user-given ratings (e.g., creativity and trust) towards their experience during the co-creation process. 
The findings lead us to believe that there exists a balance between user agency and AI initiative that can be aligned to better support co-creation. An ideal design would likely involve a dynamic adjustment of AI's role based on the user's expertise, background, and desire for control, allowing for both an engaging co-creation experience and innovative outcomes. 
Future designs should prioritize user feedback and adaptability, ensuring that the AI system can recognize and respond to user needs and preferences throughout the co-creation process.


\subsubsection{Customizing Co-Creation based on Researcher's Persona}
The findings of RQ3 revealed that the background of researchers, including factors such as domain knowledge and research experience, were found to influence their interaction with our system and their evaluation of generated RQs. Additionally, a user's preference for the diversity and specificity of the model's outputs can vary depending on their research stage.
For instance, during earlier stages of research, users tend to prefer the generated RQs to be more explorative and cover a broader perspective, while they might value more relevant and specific outputs during later stages of research where the research topic or idea has been scoped down to a certain degree. 
Although prior research has described the information-seeking behavior of researchers as a non-linear and dynamic process \cite{foster2004nonlinear}, our observations suggest that users' expectations of the system diverge based on their individual backgrounds and the progression of their research.
The newly identified factors highlight the significance of personalization and adaptability to users' evolving needs in co-creation systems, particularly in the context of scholarly research.
Recent research in aligning user persona with LLMs \cite{hwang2023aligning,salemi2023lamp} has provided viable means for future designs of personalized research co-creation systems. 
It's crucial that later designs emphasize adaptability to ensure that system outcomes align with the individual researcher's background, progress, and specific research goals.




\subsection{Design Implications}



\subsubsection{Explaining LLM Rationales Through Mind-Map-Styled Design}
Besides harnessing the text generation capability of LLMs, our study also highlighted the importance of utilizing the Chain-of-Thoughts prompting ability to not only improve LLMs' task-specific performance \cite{wei2022chain}, but also as a way to enhance the explainability of AI-based systems and bridge the gap for users to understand the rationale of AI. 
Prior research has explored using mind-map-like design \cite{kang2021metamap} to support users' creative ideation process.
Our \textit{CoQuest} system explored another viable design of using mind-map-styled visualization to facilitate a natural communication of LLMs' chain of thoughts towards humans in addition to the modality of text, most importantly through the \textit{AI Thoughts} feature.
It was also noted that some participants encountered difficulty interpreting AI rationales even after reading the explanation provided through \textit{AI Thoughts}, as they could not ask for further explanations as in common chat-based interfaces. 
We argue for future designs of AI-based co-creation systems to provide explanations of AI rationales interactively through graphical designs, such as dynamic exploration features where users can prompt for further explanations or ask questions directly within a mind-map-like interface. 


\subsubsection{Sharing of Expertise: Steering the Direction of Outputs by Injecting Meta-Research Knowledge}
The \textit{CoQuest} can not only be used for co-creation, but also for educational purposes that transfer knowledge about meta-research practices among researchers of different levels of experience. 
During our user study, we observed that while some researchers focused on the novel elements AI provided in newly generated RQs in a brainstorming manner, researchers with more experience tended to explicitly indicate their needs for AI to generate results from a more ``technical'' perspective, such as providing ideas related to evaluation metrics or surveying existing works regarding certain research methodology.
Although users often have different specific topics of interest during RQ co-creation, the higher-level research thinking and skills, as discussed in existing meta-research works \cite{weissgerber2021training,ioannidis2015meta,stevens2023using}, can be beneficial in general when shared across users, especially for novice researchers or researchers new to certain fields. 
Past research has explored designs to facilitate the cultivation of new researchers' research skills, such as storytelling \cite{schrum2022cultivating}. 
New understandings unveiled by this study about researchers' behavior using LLM-enabled co-creation to provide novel implications that can pave the way for a more collaborative and educative approach in future system designs. Experienced researchers' interaction with the co-creation system can provide pathways from which less experienced researchers can learn and benefit. Integrating this understanding, we envision a design where the AI serves not just as a tool for co-creation, but also as a mediator in the knowledge transfer process between researchers. This integration can potentially bridge the gap between research idea formulation across domains by utilizing user-sourced expertise of meta-research.

\subsubsection{Utilizing Personalized Design to Harness ``Surprising'' Outputs} 
Hallucination is a well-recognized challenge in many task-oriented LLM system designs \cite{Singhal2022LargeLM,Li2023HaluEvalAL,McKenna2023SourcesOH}. Previous HCI studies suggest that while users might sometimes view unexpected outputs favorably \cite{Epstein2022WhenHA}, they can also find them unhelpful at times \cite{Lee2022CoAuthorDA}.
Our findings echo these observations. While certain users viewed AI's topic drift-off unfavorably, others appreciated the fresh content introduced by the AI. This points to a subjective user preference. 
Additionally, our findings in RQ3 reflected that users' background influences their expectations for outputs generated by AI.
Thus, we argue that in the future design of human-AI co-creation systems, the decision to allow models to introduce potentially out-of-context content (often labeled as ``hallucination'') should be seen as a design choice rather than a blanket problem. Furthermore, user backgrounds, such as domain familiarity, should be taken into account as they could influence the optimal design options.
Future designs might consider how to detect users' different intents via their feedback. 
One possible approach could be to utilize implicit preference probing, where the system actively gauges the user's reaction to certain outputs and adjusts its responses accordingly. For instance, if a user consistently displays positive engagement with unexpected outputs, the AI could be more inclined to provide similarly ``out-of-context'' suggestions in future responses. 
Similar designs can also be applied to identify which stage of the research or creative process a user is in, so that the AI can tailor its responses.

\subsection{Ethical Concerns and Potential Biases} 

\subsubsection{\bluehighlight{Ethical Implications of LLM Usage: Plagiarism and Hallucination}}

\bluehighlight{
The use of LLMs in research idea co-creation processes can lead to potential ethical concerns, e.g., risks of plagiarism and hallucination. 
LLMs generate content based on their vast training data, which raises concerns about the originality of their outputs. When asked about their concerns over the \textit{CoQuest} system, several participants raised their concerns about the originality of ideas generated by the LLM-based backend. The possibility that an LLM might inadvertently replicate existing content without proper attribution, even when they are not directly copying content from the training corpus, poses a significant challenge to the integrity of research and creative processes \cite{gingerich2013claiming}. This is particularly relevant in academic contexts where the originality of ideas and proper citation are at core \cite{orenstrakh2023detecting}.
Additionally, LLMs are known to suffer from the issue of hallucination \cite{zhang2023siren, rawte2023survey}. This characteristic of LLMs can mislead users, especially those who might be new to a research domain. Reliance on hallucinated content could lead to erroneous conclusions or decisions especially during the earlier stages of a research lifecycle.
To mitigate these risks from the perspective of designing LLM-based co-creation systems for scientific research, it is essential to verify the credibility of LLM-generated content and apply methods to ensure proper attribution of scientific ideas from other researchers. This may be achieved through methods like fact-checking \cite{manakul2023selfcheckgpt}, grounding LLM-generated content within credible sources \cite{yue2023automatic}, and developing understanding among users to recognize potential flaws in LLM-generated content. 
Addressing these ethical concerns is crucial for maintaining the integrity and reliability of LLM-based human-AI co-creation, particularly in fields where intellectual property rights are highly valued.
}

\subsubsection{\bluehighlight{User biases and blind spots}}
In our study, we found that users' expertise influenced their sense of control when co-creating with AI. Specifically, users felt less control and lower cognitive demand when working on familiar topics. Interviews with participants revealed that they had taken cognitive shortcuts by having intrinsic expectations about the generated outcomes. Their sense of control also decreased when these expectations were not met, indicating the presence of confirmation biases. This aligns with the growing research on understanding cognitive biases in human-AI interactions \cite{boonprakong2023workshop,boonprakong2023bias,wang2019designing,DIS23si}. Our research further highlights the impact of confirmation biases on users' perceived control in human-AI co-creation tasks. Moreover, our research suggests that these biases can operate through the lens of human expertise, potentially creating blind spots. Future research should explore diverse strategies for mitigating bias in co-creation with AI, such as drawing from crowd-sourced ideas, as demonstrated in \citet{Eric2023}'s work.

\subsubsection{\bluehighlight{Over-reliance on AI}}
Another ethical concern with our system is users blindly accepting AI-generated outcomes or relying too heavily on them, as pointed out by \citet{buccinca2021trust}. To address this, we implemented two effective approaches in our design. First, we introduced an RQ rating during the generation process to encourage users to evaluate AI-generated content. Second, we incorporated AI thoughts to assist users in understanding the literature space and the generation process. Users frequently utilized AI thoughts during wait times, which enhanced their perceived control in co-creation. These two strategies, utilizing metrics to promote human active evaluation of AI-generated content and providing explanations to enhance human understanding of the AI-generation process, offer valuable insights for those seeking to mitigate bias in human-AI collaboration. Our work builds upon prior research \cite{vasconcelos2023explanations} by providing options for users to check for explanations in order to reduce AI over-reliance. 

Nonetheless, our study did not provide empirical understanding of the longer-term impact of using an LLM-based RQ co-creation system. 
Prior research \cite{loi2020societal} has pointed out that certain designs in current AI systems could hinder human creativity development in the long term.
We argue that further studies should be conducted to understand both the positive and negative impacts of human-AI co-creation systems for research ideation longitudinally over human researchers regarding aspects such as creativity, research preferences, and behaviors during ideation.
Future research should also explore more ways of explaining AI outputs and designs to promote active human thinking and advance the understanding of the role of explanations in research and learning.


\section{Limitations and Future Work}

We acknowledge that our study has several limitations. First, our system currently relies on a fixed and relatively limited set of publications as its source pool. This design limitation often results in users finding themselves constrained to a narrow segment of literature after a few iterations of RQ generation. Ideally, the system should be integrated with online publications databases to harness a broader and continuously updated spectrum of publications. 
Trust in the system also emerged as a concern. Some users hesitated to utilize the generated RQs directly with concern about their originality and fearing potential overlaps with pre-existing research. Addressing these concerns is crucial for enhancing user confidence and the overall effectiveness of the system.
Furthermore, the time constraints imposed on the tasks might not have been optimal for all participants. Those less familiar with the task might require more time before they get acquainted with the research space and can effectively generate RQs. This ``warm-up'' period could be considered more thoughtfully in future studies.
\bluehighlight{Additionally, we acknowledge the limitation in our analysis related to user behavior beyond wait times. Our system did not precisely record the completion times for each RQ generated.
Due to the extensive duration and the complexity of behaviors in these periods, these activities were not systematically coded and analyzed, which may have omitted valuable insights into user interaction with the system.}
Moreover, our evaluation focused solely on doctoral students who, while having some research background, are still in the early stages of their research careers. The results might differ when evaluating seasoned researchers or even undergraduates. 
Expanding the participant pool in future studies can offer a more comprehensive understanding of the system's effectiveness and user experience across varying expertise levels.
Future work should also study the longitudinal effect of the \textit{CoQuest} system usage over human researchers.



\section{Conclusion}
In this study, we introduced an agent LLM system, called \textit{CoQuest}, aiming to support the creation of research questions.
Through a formative study with actual researchers, we proposed a mental model combining the process of literature discovery and research ideation and applied it to the design of the \textit{CoQuest} system. 
We introduced two interaction design options for \textit{CoQuest}: breadth-first and depth-first generations, diversifying the degree of AI's initiative during co-creation. 
A within-subjective study with 20 participants revealed that a higher degree of AI initiative led to co-creation outcomes with enhanced creativity, albeit at the expense of the overall co-creation experience. 
We also found that users who effectively utilized wait time experienced higher-quality outcomes and developed a stronger sense of control.

\begin{acks}

This material is based upon work supported by the National Science Foundation under Grant No. 2119589. Any opinions, findings, and conclusions or recommendations expressed in this material are those of the author(s) and do not necessarily reflect the views of the National Science Foundation.
Additionally, results presented in this paper were obtained using CloudBank \cite{norman2021cloudbank}, which is supported by the National Science Foundation under award No. 1925001.

\end{acks}

\bibliographystyle{ACM-Reference-Format}
\bibliography{references}

\clearpage
\appendix
\onecolumn
\section{Appendices}
\label{appendix}


\subsection{Post-Session Survey Questions}
\label{apdx:survey_questions}
\begin{table}[h]
\centering
\caption{Survey Questions for User Experience Evaluation}
\label{tab:survey_questions}
\begin{tabularx}{\textwidth}{lX}
\toprule
\textbf{Aspect} & \textbf{Question} \\
\midrule
Control & How much control did you feel you had when using the system? \\
& Did the system allow you to make choices and decisions while creating RQs? \\
\addlinespace
\hline
\addlinespace
Creativity & How would you rate the creativity of the research questions generated by the system? \\
& How creative did you feel about yourself when using the system to create RQs? \\
\addlinespace
\hline
\addlinespace
Meta-Creativity & Did the system inspire new ways of thinking or approaching RQ creation for you? \\
& Did the generated RQs make you think or reflect about the topic in a new way? \\
\addlinespace
\hline
\addlinespace
Cognitive Load & How mentally demanding was the task using the system? \\
& Did you feel overloaded with information or options while using the system? \\
\addlinespace
\hline
\addlinespace
Trust & How confident were you in the RQs generated by the system? \\
& Would you trust the system's generated RQs to be used in a real research scenario? \\
\bottomrule
\end{tabularx}
\end{table}

\subsection{A Mixed-Effect Model of User Feedback Length}

\label{apdx:mixed_effect_model}

We observe that users tend to provide human feedback to the system in distinct ways. 
In terms of linguistic styles, we investigated the difference among users' input lengths (word counts) by fitting a mixed-effects model by considering both the random effect from different users (i.e., $b_{0i}$), and the potential fixed effect from the two conditions (i.e., $\beta_1$):

\begin{equation}
    Y_{ij} = \beta_0 + \beta_1 \times \text{condition} + b_{0i} + \epsilon_{ij}
\end{equation}

Where $Y_{ij}$ is the length of text feedback for the $j^{th}$ observation of the $i^{th}$ participant;
$\beta_0$ is the intercept;
$\beta_1$ is the effect of the two conditions (i.e., depth-first and breadth-first);
$\text{condition}$ is a categorical variable (binary) indicating the user's condition;
$b_{0i}$ is the random effect for the \( i^{th} \) user;
$\epsilon_{ij}$ is the residual error for the $j^{th}$ observation of the $i^{th}$ user.

\begin{table}[h!]
    \centering
    \begin{tabular}{lcccccc}
        \toprule
        & Coef. & Std.Err. & z-value & \( P(>|z|) \) & [0.025 & 0.975] \\
        \midrule
        Intercept & 10.119 & 0.908 & 11.147 & 0.000 & 8.339 & 11.898 \\
        condition (depth-first=1) & -0.637 & 1.034 & -0.616 & 0.538 & -2.663 & 1.389 \\
        \midrule
        Group Var & 4.411 & 0.405 & & & & \\
        \bottomrule
    \end{tabular}
    \caption{Mixed linear model regression results for feedback lengths}
    \label{tab:mixed_model_results}
\end{table}

The results of the mixed-effect model are shown in Table \ref{tab:mixed_model_results}.

\subsection{
\bluehighlight{
Examples of LLM-based Agent Prompts and Responses
}
}

\label{apdx:prompt_examples}

The CoQuest system is implemented based on the similar prompting method from AutoGPT. At each step of inference, the model generates the response with the next action to take. The action is then executed by the system, with the returned response appended to the input for the next step of inference. 
The detailed prompts used in the system are as follows.

\subsubsection{System prompt}
The system prompt is designed to indicate the overall tasks and constraints for the LLM agent.

\newpage

\vspace{1cm}
\textbf{System prompts:}
\begin{lstlisting}
You are research-GPT, an AI agent designed to automate the creation process of research questions/ideas, literature survey, and brainstorming.

Your decisions must always be made independently without seeking user assistance. Play to your strengths as an LLM and pursue simple strategies with no legal complications.

GOALS:
1. Survey relevant past research papers/works
2. Summarize these works into novel findings and insights
3. Come up with novel research questions, and also generate possible expected results for each research question
4. Summarize the novelty and similarity between your proposed new research questions, and past research
5. Further survey literatures, and refine the research questions


Constraints:
1. No user assistance
2. Exclusively use the commands listed in double quotes e.g. "command name"
3. Use subprocesses for commands that will not terminate within a few minutes
4. First collect paper information, and then generate RQs.Do not create Agents to collect information, only rely on the query command.

Commands:
1. Summarize Existing Papers: "search_and_summarize_papers", args: "query": "<text>"
2. Hypothesizing Use Cases: "hypothesize_use_cases", args: "context": "<text>"
3. Narrow down RQs: "narrow_down_rqs", args: "context": "<text>"
4. Comparing existing RQ with existing papers: "compare_rq_with_papers", args: "past_research_summary": "<text>", "rqs": "<text>"

Performance Evaluation:
1. Continuously review and analyze your actions to ensure you are performing to the best of your abilities.
2. Constructively self-criticize your big-picture behavior constantly.
3. Reflect on past decisions and strategies to refine your approach.
4. Every command has a cost, so be smart and efficient. Aim to complete tasks in the least number of steps.

You should only respond in JSON format as described below.
Response Format: 
{
"thoughts": {
       "text": "thought",
       "reasoning": "reasoning",
        "plan": "- short bulleted\\n- list that conveys\\n- long-term plan",
        "criticism": "constructive self-criticism",
        "speak": "thoughts summary to say to user"\n    },
    "command": {
        "name": "command name",
        "args": {
            "arg name": "value"
        }    },
    "RQs": {
        "rq1": "{ACTUAL_RQ}",
        "rq2": "{ACTUAL_RQ}",
\end{lstlisting}

\begin{lstlisting}
        "rq3": "{ACTUAL_RQ}"
    }} 
Ensure the response can be parsed by Python json.loads. Each response should contain 3 research questions (RQs) proposed based on the current inputin the JSON format specified above, by replacing the ACTUAL_RQ with your RQ. The key RQs should also be on the top level of the json object.  You should always generate valid and non-empty RQs, even if the input is not clear enough.  Always be constructive, specific, and creative. 
\end{lstlisting}

\subsubsection{Prompts used for each action}
We design each action based on the mental model as an individual task to be completed by the agent. For each action, a separate function is invoked to trigger a new turn of LLM inference. 

\vspace{1cm}
\textbf{Search and summarize papers:}
\begin{lstlisting}
Summarize the input literatures into 5 bullet points. 
Always explain each point in detail and assume the user has no background knowledge.
Your reply should strictly be in the following format in one line:
    Here is a summary of some existing works:
        1. ...
        2. ...
        3. ...
\end{lstlisting}

\vspace{1cm}
\textbf{Hypothesize use cases:}
\begin{lstlisting}
You are a helpful AI that can hypothesize use cases for users.
I will provide you with some context, and you should generate three use cases based on the context.
Your reply should strictly be in the following format in one line:
    Here are some potential use cases based on the current RQ:
        Use case 1: ...
        Use case 2: ...
        Use case 3: ...
\end{lstlisting}

\vspace{1cm}
\textbf{Narrow down RQs:}
\begin{lstlisting}
You are a helpful AI that can narrow down RQs for users.
I will provide you with some context, and you should reflect and generate a list of bullet points that narrows down the context.
The reply should be a list of bullet points, each bullet point should be a sentence:
    To narrow down the RQ, we should consider the following:
        - ...
        - ...
        - ...
\end{lstlisting}

\newpage

\vspace{1cm}
\textbf{Compare RQ with papers:}
\begin{lstlisting}
You are a helpful AI that can compare RQs with existing papers for users.
I will provide you with some context, and you compare the RQs with existing papers, and provide a summary of the findings.
The reply should be a list of bullet points, each bullet point should be a sentence:
    Here are some findings from comparing the RQs with existing papers:
        - ...
        - ...
        - ...
\end{lstlisting}

\subsubsection{Triggering prompt}
The triggering prompt is appended to the end at each inference, to further control the generation.

\vspace{1cm}
\textbf{Triggering prompt:}
\begin{lstlisting}
Determine which next command to use, and respond using the format specified above. You should always revise your old RQs into new RQs, based on the previous context and user input. Be specific and creative.Always go deeper on the high level RQ, do not repeat RQs that are already in history. Remember, always generate RQs in the format specified above by replacing the ACTUAL_RQ with real RQs. 
\end{lstlisting}

\end{document}